# ECOSoundSet: a finely annotated dataset for the automated acoustic identification of Orthoptera and Cicadidae in North, Central and temperate Western Europe

David Funosas[1 2 3], Elodie Massol[2 3], Yves Bas[4 5], Svenja Schmidt[6], Dominik Arend[6], Alexander Gebhard[7], Luc Barbaro[8], Sebastian König[7], Rafael Carbonell Font[9], David Sannier, Fernand Deroussen[10], Jérôme Sueur[11], Christian Roesti[12], Tomi Trilar[13], Wolfgang Forstmeier[14], Lucas Roger[15 16], Eloïsa Matheu[17], Piotr Guzik[18], Julien Barataud[19], Laurent Pelozuelo[3], Stéphane Puissant[20], Sandra Mueller[6], Björn Schuller[6 21], Jose M. Montoya[1], Andreas Triantafyllopoulos[7], Maxime Cauchoix[1]

[1]Station d'Écologie Théorique et Expérimentale (SETE, CNRS), Moulis, France

[2]Université Paul Sabatier - Toulouse III, UPS, Toulouse, France

[3]Centre de Recherche sur la Biodiversité et l'Environnement - UMR 5300 CNRS-INPT-IRD-UT, Toulouse, France

[4]Centre d'Ecologie et des Sciences de la Conservation (CESCO, MNHN), Centre National de la Recherche Scientifique, Sorbonne Université, Paris, France

[5]PatriNat (OFB, MNHN), 75005 Paris, France

[6]University of Freiburg, Faculty of Biology, Geobotany, Schaenzlestr. 1, D-79104 Freiburg, Germany

[7]CHI – Chair of Health Informatics, MRI, Technical University of Munich, Germany

[8]Dynafor, INRAE-INPT, University of Toulouse, Castanet-Tolosan, France

[9]Institució Catalana d'Història Natural (ICHN), Barcelona, Spain

[10]Nashvert Naturophonia, Val Maravel, France

[11]Institut de Systématique, Evolution, Biodiversité (ISYEB), Muséum national d'Histoire Naturelle (MNHN), CNRS, Sorbonne Université, Ecole Pratique des Hautes Etudes - PSL, Université des Antilles, Paris, France

[12]Orthoptera.ch, Bern, Switzerland

[13]Slovenian Museum of Natural History (PMSL), Ljubljana, Slovenia

[14]Department of Ornithology, Max Planck Institute for Biological Intelligence, Seewiesen, Germany

[15]INRAE, Université de Bordeaux, BIOGECO, Pessac, France

[16]Plante & Cité, Angers, France

[17]Museu de Ciències Naturals de Barcelona (MCNB), Barcelona, Spain

[18]Murowaniec 44, 38-455 Niżna Łąka, Poland

[19]117 rue Jean Carou - 19330 Chanteix

[20]Muséum d'Histoire Naturelle, Dijon, France

[21]Group on Language, Audio, & Music (GLAM), Imperial College London, UK

# Abstract

## Background

Recent studies suggest a widespread and substantial decline in insect abundance and diversity across European terrestrial ecosystems. This entails an urgent need for effective large-scale insect monitoring methods to determine the extent of the problem and to understand the global and local mechanisms driving this decline. Passive acoustic monitoring (PAM) enables the monitoring of sound-producing insect populations and communities at an unprecedented temporal and spatial scale by remotely capturing sounds such as stridulations, timbalizations and wingbeats. However, currently available tools for the automated acoustic recognition of European insects in natural soundscapes are limited in scope. Hence, the development of algorithms capable of reliably identifying a broad range of European insect sounds will greatly enhance the ability of PAM to meaningfully assist in the characterization of sound-producing insect communities, especially orthopterans and cicadas. Large and ecologically heterogeneous acoustic datasets are currently needed for these algorithms to cross-contextually recognize the subtle and complex acoustic signatures produced by each species, thus making the availability of such datasets a key requisite for their development.

## Methods

Here we present ECOSoundSet (European Cicadidae and Orthoptera Sound dataSet), a dataset containing 10,653 recordings of 200 orthopteran and 24 cicada species (217 and 26 respective taxa when including subspecies) present in North, Central, and temperate Western Europe (Andorra, Belgium, Denmark, mainland France and Corsica, Germany, Ireland, Luxembourg, Monaco, Netherlands, United Kingdom, Switzerland), collected partly through targeted fieldwork in South France and Catalonia and partly through contributions from various European entomologists. The dataset is composed of a combination of coarsely labeled recordings, for which we can only infer the presence, at some point, of their target species (weak labeling), and finely annotated recordings, for which we know the specific time and frequency range of each insect sound present in the recording (strong labeling). We also provide a train/validation/test split of the strongly labeled recordings, with respective approximate proportions of 0.8, 0.1 and 0.1, in order to facilitate their incorporation in the training and evaluation of deep learning algorithms.

**Conclusions**

This dataset could serve as a meaningful complement to recordings already available online for the training of deep learning algorithms for the acoustic classification of orthopterans and cicadas in North, Central, and temperate Western Europe.

## 1. Introduction

The substantial and widespread decline in terrestrial European insect populations suggested by recent studies (Conrad et al., 2006; Goulson et al., 2008; Thomas et al., 2016; Hallmann et al., 2017; Forister et al., 2019; Seibold et al., 2019; Pilotto et al., 2020; van Klink et al., 2020; Fox et al., 2021; van Klink et al., 2024) raises profound ecological concerns. Long-term monitoring data reveal sharp reductions in both abundance and richness (van Strien et al., 2019; Widmer et al., 2019; Dirzo et al., 2014; Møller, 2020), with some regions reporting losses exceeding 75% of total flying insect biomass over the past few decades (Hallmann et al., 2017). This decline has been attributed to a confluence of anthropogenic factors such as habitat loss, agricultural intensification, pesticide use, light pollution and climate change, which together could be leading to a "death by a thousand cuts" (Wagner et al., 2021; Rumohr et al., 2023). This plurality of ecological stressors, coupled with the current paucity of long-term insect population data in Europe (Eisenhauer et al., 2019; van Klink et al., 2021;

Rumohr et al., 2023), underscores an urgent need to develop effective methods to monitor insect populations at large temporal and spatial scales. Such methods are crucial for better understanding the global and local mechanisms driving these losses and for devising well-targeted conservation strategies.

Passive acoustic monitoring (PAM) appears as a promising method to improve our understanding of these trends by providing a scalable, highly standardized, cost-effective, non-lethal and non-invasive way of obtaining species distribution data for sound-producing animals (Darras et al., 2018; Darras et al., 2019; Melo et al., 2021; Napier, 2024). Despite having mostly been used for the study of vertebrates such as birds (Bobay et al., 2018; Barbaro et al., 2023; Brunk et al., 2023; Bielski et al., 2024), bats (Claireau et al., 2019; Hoggatt et al., 2024), and anurans (Melo et al., 2021; Chen et al., 2023; Bota et al., 2024), recent studies suggest that PAM could also serve as a powerful tool to monitor insect populations by capturing sounds such as orthopteran stridulations (Newson et al., 2017; Riede et al., 2024; Symes et al., 2024; Thibault et al., 2024), cicada timbalizations (Gasc et al., 2018; Do Nascimento et al., 2024; Attinger et al., 2025), and wingbeats (Rodríguez Ballesteros et al., 2024). Data collected through PAM can ideally complement on-site active insect surveys by improving the detectability of species whose acoustic activity patterns do not coincide with the dates and times at which active monitoring is usually conducted (Sebastián-González et al., 2018), and by allowing to study the variations in acoustic activity patterns along a given time gradient across a large number of replicates (Gasc et al., 2013; Towsey et al., 2014).

Even though PAM allows for the efficient collection of large amounts of ecoacoustic data, the ability to process and analyze the resulting acoustic datasets in a reliable and scalable manner remains a significant bottleneck, particularly for the study of insects. Unlike birds or bats, for which algorithms such as BirdNET (Kahl et al., 2021) and Tadarida (Bas et al., 2017) have revolutionized automated species identification, comparable tools for insect acoustic recognition in Europe are more limited in scope —35 grasshopper species in CrickIt (Aquila Ecology, 2024), 1 cicada species in Cicada Hunt (Rogers, 2018) and 5 grasshopper, 75 katydid and 6 cicada species in Tadarida, compared with the 222 soniferous orthopteran and 24 cicada species (245 and 26 respective taxa including subspecies) present in North, Central, and temperate Western Europe (Table S1)—. Hence, the development of algorithms capable of reliably identifying a broad range of European insect sounds could greatly reduce the current need for time-intensive manual analysis of passively collected recordings, thus enhancing the ability of PAM to meaningfully assist in the characterization of sound-producing insect communities and the assessment of long-term population trends.

The development of reliable Deep Learning (DL) algorithms for acoustic species identification relies heavily on the availability of large and heterogeneous datasets, i.e., covering a broad spectrum of environmental conditions, recording equipment, background noise profiles, and species-specific variations in sound production across geographical regions, behavioral contexts, temperature gradients and seasonal or diel cycles. Ensuring such diversity in training data is essential for improving the generalizability of automated recognition systems and mitigating biases that could arise from overfitting to narrow or context-dependent acoustic patterns. Consequently, the accessibility of comprehensive datasets is a fundamental prerequisite for enabling these models to robustly identify the subtle and complex acoustic signatures characteristic of each species across diverse contexts.

Some such datasets for European orthopterans and cicadas already exist (Faiß, 2023; Faiß et al., 2025), and an ample amount of recordings from both taxonomic groups can be freely downloaded from online libraries such as Xeno-canto, iNaturalist, observation.org, ZFMK, MinIO and BioAcoustica (Table S1). However, in these datasets, each audio file is labeled after a single species despite the potential acoustic presence of multiple other species in the recording background. This means that, in some moments over the duration of a given recording, non-target species could be emitting sounds in the absence of the labeled species, potentially confusing the algorithm being trained on these recordings and resulting in a suboptimal recognition of the acoustic signature of each species. In contrast to such weakly labeled recordings, strongly labeled ones provide precise temporal and spectral coordinates of the target signal within the spectrogram, specifying its time and frequency range. Recent studies seem to indicate that DL algorithms trained with a combination of both weakly and strongly labeled material perform better than algorithms trained with weakly labeled material alone (Hershey et al., 2021; Otálora et al., 2021; Das et al., 2023). This suggests that adding a set of strongly labeled insect recordings to the collection of weakly labeled recordings already available online could enable the training of DL algorithms with greater predictive power.

Here, we present the ECOSoundSet (European Cicadidae and Orthoptera Sound dataSet), an acoustic dataset with recordings of 200 orthopteran and 24 cicada species (217 and 26 respective taxa when including subspecies) present in North, Central and temperate Western Europe (Andorra, Belgium, Denmark, mainland France and Corsica, Germany,

Ireland, Luxembourg, Monaco, Netherlands, United Kingdom, Switzerland), collected in part through targeted fieldwork in South France and Catalonia and in part through contributions from many European entomologists, bioacousticians and ecoacousticians. While primarily focused on the aforementioned regions, ECOSoundSet also covers a substantial proportion of species found in other parts of Europe (Fig. 1). The dataset is composed of a combination of weakly labeled recordings, for which we can only infer the presence, at some point, of their target species within the spectrogram, and strongly labeled recordings, for which we know the exact stridulation or timbalization times of each species present in the recording. We also provide a train/validation/test split of the strongly labeled recordings, with respective proportions of 0.8, 0.1 and 0.1, in order to facilitate their incorporation in the training and evaluation of DL algorithms for the acoustic classification of orthopterans and cicadas. To the best of our knowledge, this is the first publicly available dataset to cover the majority of soniferous orthopteran and cicada species within a defined biogeographic region, a key factor for enabling the ecological operationality of the associated DL algorithm.

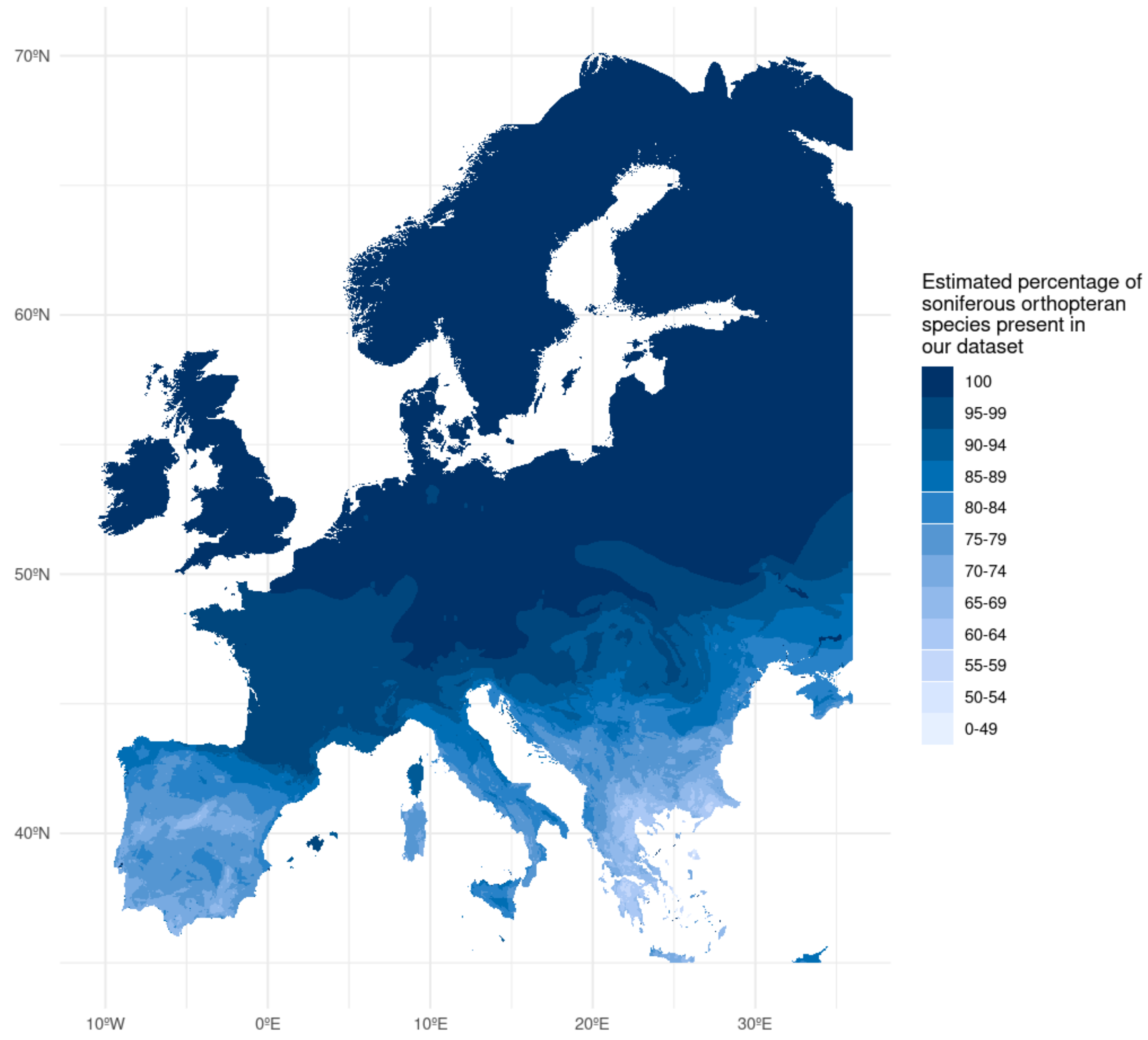

**Figure 1:** Estimated proportion of soniferous orthopteran species covered by our acoustic dataset across Europe. The orthopteran species considered to be soniferous are all those for which at least one recording has been uploaded to an online repository (GBIF.org, 2025b). Species distribution data were obtained from the International Union for Conservation of Nature (IUCN, 2016).

# 2. Materials and methods

## 2.1. Data collection

Our acoustic dataset comprises four categories of recordings: expressly collected focal recordings, expressly collected soundscapes, pre-existing focal recordings, and pre-existing soundscapes (14%, 3%, 69% and 14% of recordings, respectively; Table 1). Expressly collected focal recordings were obtained by deliberately seeking and recording orthopteran and cicada species, especially targeting those with a particular paucity of acoustic data available online due to their limited distribution range or their low rate of acoustic activity. These recordings were made, for the most part, with a Zoom H4n recorder and its built-in microphone. Expressly collected soundscapes were obtained by installing different versions of AudioMoth (Hill et al., 2018) and Song Meter recorders (Wildlife Acoustics) in the field, and both pre-existing focal recordings and soundscapes were obtained by contacting various European entomologists, bioacousticians and ecoacousticians who granted us permission to incorporate their recordings into our dataset (Table S2). Metadata for contextual variables such as the recording date, time, country, region, municipality and geographic coordinates, as well as weather conditions and air and substrate temperatures at the time of recording, are available for 88%, 54%, 91%, 88%, 75%, 44%, 13%, 22%, and 22% of recordings, respectively. Automatically calculated metadata corresponding to the main acoustic parameters of each recording, including sampling frequency, bit rate, number of audio channels and total recording duration, are provided for all recordings.

| Recording category | Number of recordings | Number of exhaustively annotated recordings | Number of partially annotated recordings | Total minutes recorded | Number of 4-second audio segments annotated |
|---|---|---|---|---|---|
| **Expressly collected focal recording** | 1469 | 195 | 1083 | 669 | 7038 |
| **Borrowed focal recording** | 7406 | 526 | 308 | 8318 | 5671 |
| **Expressly collected soundscape** | 288 | 3 | 285 | 90 | 1085 |
| **Borrowed soundscape** | 1490 | 0 | 1490 | 3142 | 15818 |

**Table 1:** Numerical overview of the data contained in the acoustic dataset for each recording category

In addition to the acoustic dataset presented in this publication, a CSV file containing the metadata and download links for a selection of publicly available recordings is also provided in the Zenodo repository (https://doi.org/10.5281/zenodo.15043893). This file includes all recordings of orthopteran and cicada species from North, Central, and temperate Western Europe uploaded before February 24, 2025, on Xeno-canto, iNaturalist, observation.org, ZFMK, MinIO and BioAcoustica. The following recordings, however, were filtered out from the list: 1) heterodyne recordings, due to the impossibility of retrieving insect sounds in their original frequencies; 2) recordings without any license attached, due to the impossibility of using them without the explicit permission of their authors; and 3) recordings lacking research grade status (i.e., without consensus from at least two users on species identification) on iNaturalist, discarded in order to minimize misidentifications. In recordings lacking subspecies-level identification, the subspecies was inferred based on the recording coordinates and the known distribution of each subspecies found in North, Central, and temperate Western Europe (Cigliano et al., 2025). This inference was performed only when a single subspecies is known to occur in the recorded country in order to prevent identifications of dubious accuracy. Recordings in time expansion were included after being converted back to their original sampling frequency and speed. The final selection of online recordings represents a total of 21,869 recordings, covering 200 orthopteran and 22 cicada species (208 and 22 respective taxa when including subspecies).

A GitHub repository has also been created to enable users to automatically retrieve the same dataset (https://github.com/DavidFunosas/GBIF_recording_download). The repository includes a script to download all recordings from Xeno-canto, iNaturalist, observation.org, ZFMK and MinIO and to extract the corresponding metadata based on GBIF results (GBIF.org, 2025a). To avoid duplicate downloads, the script ensures that recordings are not redownloaded if an entry from the same species, date, and author has already been retrieved from another platform.

## 2.2. Audio annotation

Due to the contribution of multiple research teams to the annotation of recordings, the audio annotation process was conducted following two different protocols. The first protocol, comprising the vast majority (85%) of recordings, consisted in annotating recordings with the sound edition software Audacity by drawing and labeling time-frequency bounding boxes around sounds on recording mel-scale spectrograms (Fig. S1a). As a general rule, multiple iterations of a given sound by the same individual were annotated under a single bounding box provided that the separation between consecutive iterations did not exceed 1 second. In case of longer separations, each sound iteration was annotated individually, with time-frequency bounding boxes fitting tightly to the target sound. Recordings with katydid ultrasounds were slowed down by a factor of 10 and analyzed with BatSound V4.7 in order to identify and annotate all sounds following the most up-to-date acoustic identification key for French Tettigoniidae species by Julien Barataud (Barataud, 2021a; Barataud, 2021b). These recordings, along with the corresponding annotations, were subsequently reverted to their original frequencies prior to their incorporation into our dataset.

Regarding the comprehensiveness of the annotation process, some of our recordings were exhaustively annotated, with every single sound —either biotic, anthropogenic or abiotic of natural origin (e.g. wind, rain)— being annotated, whereas other recordings were only partially annotated (see the *Data description* section for more details), with bounding boxes being drawn only around sounds of interest. All annotations were assigned a binary confidence score (1 for certain identifications and 0 for uncertain ones), which was used to filter out uncertain annotations from the final dataset. This filtering procedure implies that, even in exhaustively annotated recordings, some sounds might remain unlabeled in case of being distant or noisy enough to prevent their identification with full certainty.

The second annotation protocol also consisted in annotating sounds by tightly encapsulating them in time-frequency bounding boxes on recording mel-scale spectrograms, in this case with Raven Lite 2.0.5 (Fig. S1b). This protocol was exclusively used for the annotation of orthopteran male calls, ignoring courtship and rivalry songs as well as sounds from other taxonomic groups. Labeling varied based on stridulation style: some labeling boxes encompass multiple long echeme sequences, others contain single echemes and some only capture single syllables. Only audible sounds were annotated using this protocol.

For the labeling of biotic sound events, the Inventaire National du Patrimoine Naturel (INPN) taxonomic repository of fauna of Mainland and Overseas France (TaxRef v17.0) was used as the reference nomenclature for scientific names. All annotations and recordings in our dataset, regardless of the annotation protocol followed, were labeled to the finest achievable taxonomic resolution, including subspecies where identifiable. For the labeling of abiotic and anthropogenic sound events, a list of ad hoc sound categories was created (Table S3).

The considerable number of recordings in our dataset, combined with the time-extensive nature of the manual annotation process, meant that only a relatively modest subset of recordings (33%) could be annotated. The annotation process was conducted primarily by an expert orthopterologist (coauthor EM), with non-trivial contributions from coauthors SS, YB and DF (1967, 579, 567, and 249 recordings annotated, respectively) as well as different collaborators and research interns (see Acknowledgements).

## 2.3. Data preprocessing

The duration heterogeneity in our annotations, ranging from a few milliseconds for ultrasound-emitting katydids to several minutes for cicadas with prolonged continuous songs, posed a challenge for creating a standardized dataset suitable for training DL algorithms. To address this, each annotated recording was divided into independent 4-second segments, generating a set of spectrogram images where either some or all —in partially and exhaustively annotated recordings, respectively— of the species present are known.

The segment duration was determined through comparative trials with 3-, 4-, and 5-second segments, where the 4-second duration achieved the highest preliminary Macro F1-score. Specifically, we evaluated performance using a Convolutional Neural Network (CNN10; Kong et al., 2019) pretrained on AudioSet (Gemmeke et al., 2017) and fine-tuned on our annotated orthopteran and cicada sounds. This model can be freely accessed and used at https://huggingface.co/AlexanderGbd/insects-base-cnn10-96k-t. All audio recordings were resampled to 96 kHz, filtering out frequencies above 48 kHz in recordings with higher original sampling rates and inferring missing frequencies up to 48 kHz using a bandlimited sinc-based method (windowed sinc interpolation with low-pass filtering) in recordings with lower original sampling rates. The resulting recordings were then converted into log-mel spectrograms, and model training and evaluation were conducted using 86 species with at least 50 annotated segments. The preliminary results for 3-, 4-, and 5-second segments were almost identical, with respective F1-scores of 0.565, 0.568 and 0.566 on the independent test set (see the split procedure below). We hypothesize that the high similarity in performance across segment durations may result from an existing tradeoff between capturing the complete acoustic signature of target insect species and minimizing incidental non-target sounds within each segment.

In exhaustively annotated recordings, all detected sounds were identified and labeled, resulting in numerous audio segments containing annotations unrelated to orthopteran stridulations or cicada timbalizations. These annotations, which include birds, anurans, bats, anthropophony, and geophony, were retained in the dataset as long as they co-occurred in an audio segment with at least one orthopteran or cicada species. If an annotation spanned two adjacent audio segments, the species was marked as present in both segments provided that the portion in each exceeded a threshold of 250 ms for species producing audible sounds or 50 ms for katydids stridulating within the ultrasound range (>20 kHz). This ensured that species were not marked as present in segments where their presence was too residual to allow for a proper identification.

For the train/validation/test split, all audio segments from a given recording date and site were assigned to the same set to prevent overfitting, thus ensuring strict temporal and spatial independence between sets. Additionally, when few exhaustively annotated audio segments were available due to species-specific recording scarcity, we preferably assigned them to the test set in order to prevent the erroneous detection of False Positives and the oversight of False Negatives. These constraints resulted in many species exhibiting substantial deviations from the target respective proportions of 0.8, 0.1 and 0.1 for the train, validation and test sets (Table S1). Future dataset updates, incorporating a larger number of annotated recording sites and dates, will enhance flexibility in annotation distribution and presumptively reduce these deviations. All annotations have been transformed to and are made available in CSV format.

## 3. Data description

Our dataset comprises a total of 10,653 audio files —8,875 focal recordings and 1,778 soundscapes— with an average duration of 69 seconds and a large dispersion (SD = 167 seconds), corresponding to 204 hours of recording and a size of 129 Gb (Table 1). Recordings were collected by 82 recordists (Table S2) using 18 different sampling frequencies, with 48 kHz (34%), 44.1 kHz (22%), 96 kHz (20%), 32 kHz (9%), and 384 kHz (9%) being the most common. 6% of recordings were exhaustively annotated and 27% were partially annotated, resulting in 29,687 4-second annotation chunks (Table 1). 200 orthopteran and 24 cicada species (217 and 26 respective taxa when including subspecies) are represented, covering 90% of orthopteran and 100% of cicada species (89% and 100% of subspecies) known to make sound in North, Central and temperate Western Europe (Fig. 1, Table 2).

| **Category** | **Family** | **Number of soniferous species present in the target area** | **Number of soniferous species recorded** | **Number of soniferous species annotated** | **Number of recordings** | **Number of recordings annotated** | **Number of 4-second audio segments annotated** |
|---|---|---|---|---|---|---|---|
| **Hemiptera** | **Cicadidae** | 24 (26) | 24 (26) | 17 (17) | 2576 | 517 | 3763 |
| **Orthoptera** | **Acrididae** | 96 (110) | 84 (94) | 64 (72) | 3062 | 1114 | 4980 |
| **Orthoptera** | **Gryllidae** | 12 (12) | 12 (12) | 12 (12) | 1338 | 614 | 8511 |
| **Orthoptera** | **Gryllotalpidae** | 3 (3) | 2 (2) | 2 (2) | 115 | 24 | 282 |
| **Orthoptera** | **Tettigoniidae** | 107 (116) | 98 (105) | 64 (68) | 5608 | 2917 | 19017 |
| **Orthoptera** | **Trigonidiidae** | 4 (4) | 4 (4) | 4 (4) | 380 | 159 | 1472 |
| **Other biophony** | **-** | - | 257 | 150 | 7313 | 1487 | 3946 |
| **Anthropophony** | **-** | - | - | - | 1540 | 1540 | 4716 |
| **Geophony** | **-** | - | - | - | 418 | 418 | 1484 |

**Table 2:** Numerical overview of the data contained in the acoustic dataset for each taxonomic group. The numbers in parentheses indicate values corresponding to subspecies.

## 4. Discussion

InsectSet459, the largest and most comprehensive open dataset of orthopteran and cicada sounds published to date (Faiß et al., 2025),

comprises over 26,000 recordings from 459 different species distributed globally. Sourced from three major online platforms —Xeno-canto, iNaturalist, and BioAcoustica—, this collection represents a landmark resource for the advancement of acoustic research on these taxa. In this work, we aim to contribute a complementary dataset with several additional features: (1) over 10,000 previously unpublished orthopteran and cicada recordings, (2) fine-grained annotations (strong labeling) for 3,890 of these recordings, and (3) an open-source R script designed to streamline the automated download of recordings from Xeno-canto, iNaturalist, Observation.org, ZFMK, and MinIO and the extraction of the corresponding metadata based on GBIF results (GBIF.org, 2025a). This script will enable potential users to download all recordings uploaded and validated up to the date of use, and to limit the search to the taxonomic groups and regions of interest through the filters available on the GBIF platform.

Due to the limited availability of recordings for species endemic to the Iberian, Italian and Balkan peninsulas, we restricted the spatial extent of our dataset to North, Central and temperate Western Europe, a region where soniferous species coverage (90% of orthopteran and 100% of cicada species) is sufficient to support the training of ecologically operational DL algorithms. However, the spatial distribution of our recordings presents a considerable degree of concentration around South France and Catalonia, where most of our targeted fieldwork took place (Fig. 2). This uneven spatial distribution may lead to the algorithm overfitting to the local acoustic ecotypes (Ferguson, 2002; Pinto-Juma et al., 2005; Ivković et al., 2022; Kovalchuk, 2024; Sebastián-González et al., 2025) of the target species in the most heavily sampled regions, potentially undermining its ability to recognize acoustic ecotypes from more distant areas. That said, our labeling of insect sounds at the subspecies level where identifiable could help mitigate this issue. We also suggest caution when interpreting the spatial applicability of our dataset presented in Fig. 1, since the species-specific distribution maps upon which the figure is based (IUCN, 2016) might be partially incomplete due to knowledge gaps, especially in regions of high species richness.

The number of recordists (Table S2) and the variety of devices (≥21 recorders and ≥23 microphones) and acoustic parameters (e.g., sampling frequency, sound amplification) having been used for the collection of acoustic data may enhance the ability of DL algorithms trained on our dataset to generalize across different contexts (Ryu et al., 2024). However, this heterogeneity may also hinder the recognition of high-frequency insect sounds, whose capture varies in completeness depending on the sampling frequency used. In our dataset, the sampling frequency selected for each focal recording was generally adapted to the target species, and most nighttime soundscapes were recorded at sampling frequencies high enough to capture ultrasonic stridulations in their entirety. Nonetheless, a non-negligible portion of our soundscapes were recorded at a sampling frequency of 48 kHz to optimize battery life (see recording_metadata.csv in the Zenodo repository). While this sampling frequency covers the full frequency range of cicadas, grasshoppers, crickets, and most katydids, it resulted in some recordings presenting only a partial capture of the stridulations of ultrasound-emitting katydids.

Another challenge for the development of DL algorithms for the automatic identification of biological sounds in natural soundscapes obtained through PAM is the substantial signal-to-noise disparity between the focal recordings typically used to train the algorithms and the soundscapes they are frequently used on once developed (Fig. S2). Since individuals in focal recordings are often in close proximity to the microphone, this may hinder the ability of DL algorithms to generalize effectively to the more distant and potentially overlapped sounds commonly found in soundscape recordings. In this context, the inclusion of annotated soundscapes in our dataset could help bridge the saliency gap between insect sounds in focal recordings and those in soundscape recordings, thereby enhancing the ability of the algorithms to recognize insect sounds in the type of recordings they are most likely to used on (Liu et al., 2022).

A)

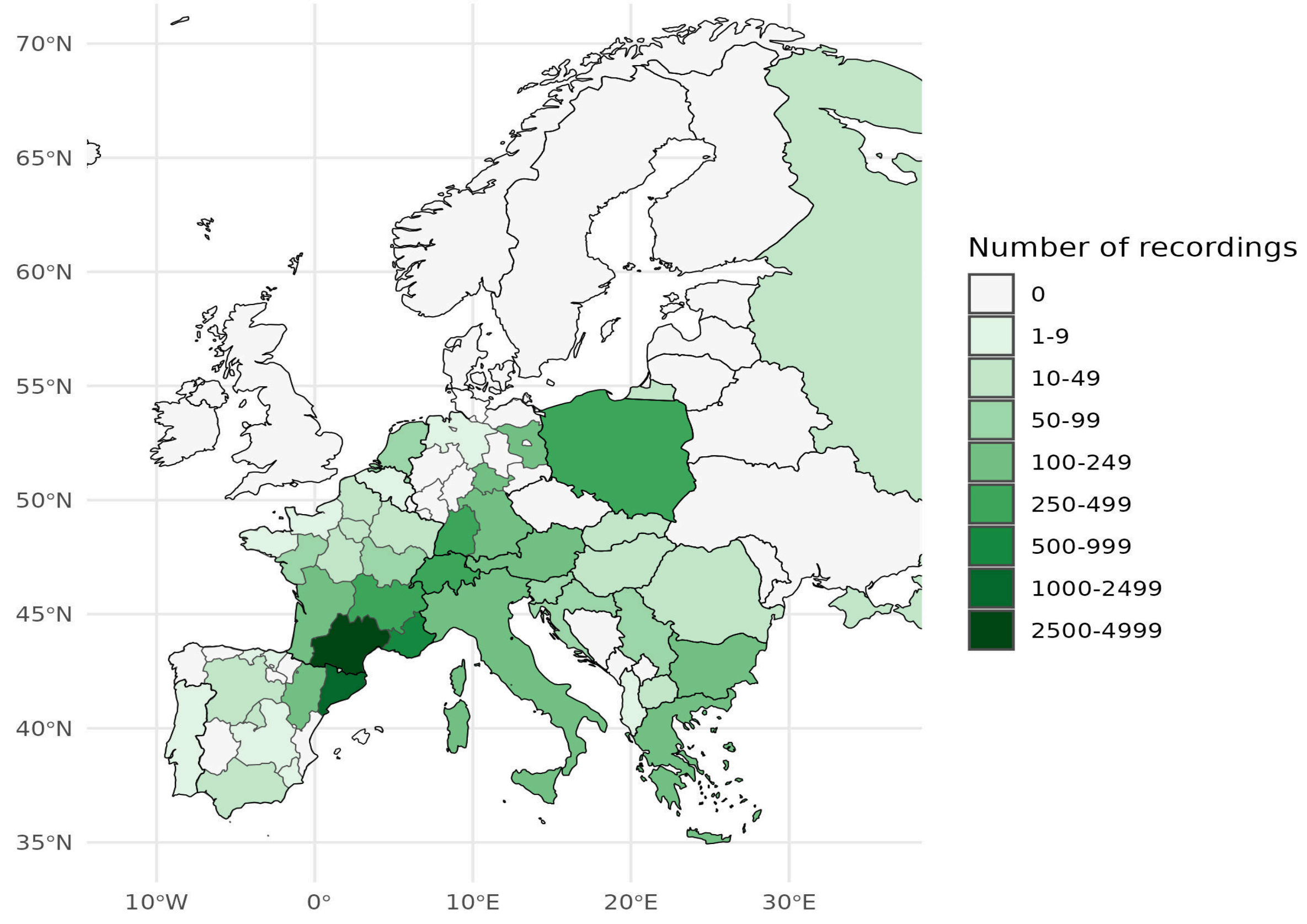

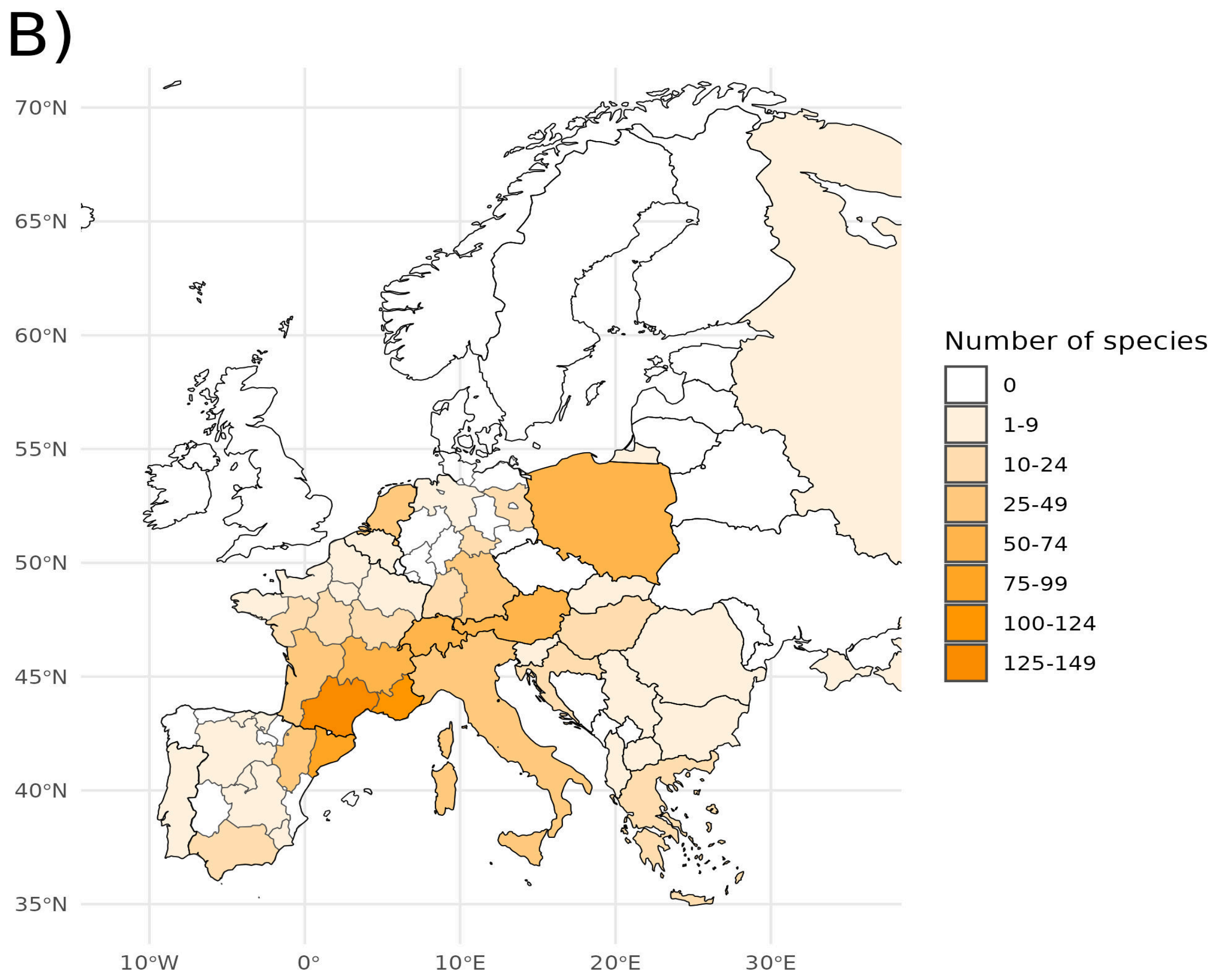

**Figure 2:** Geographical distribution of the recordings comprising the acoustic dataset. Recordings (A), as well as the number of species recorded (B), are grouped at the regional level for France, Germany, and Spain —the countries with the most recordings— and at the country level elsewhere.

Due to their modest size, our finely annotated recordings do not stand on their own as a complete dataset for the training of DL algorithms. However, they could serve as a meaningful complement to recordings already available online by providing the algorithms with the specific temporal and frequency coordinates of each insect sound within a spectrogram, thus enhancing their ability to recognize the unique acoustic signature of each species. Likewise, the provision of exhaustively annotated recordings could be particularly helpful in preventing the erroneous detection of False Positives and the oversight of False Negatives. In addition, the relatively low annotation data imbalance across species in our dataset (Table S1, Fig. 3) could partially offset the much greater imbalance in the cross-species availability of online recordings, thereby mitigating the overrepresentation of common species in the training phase of the algorithms.

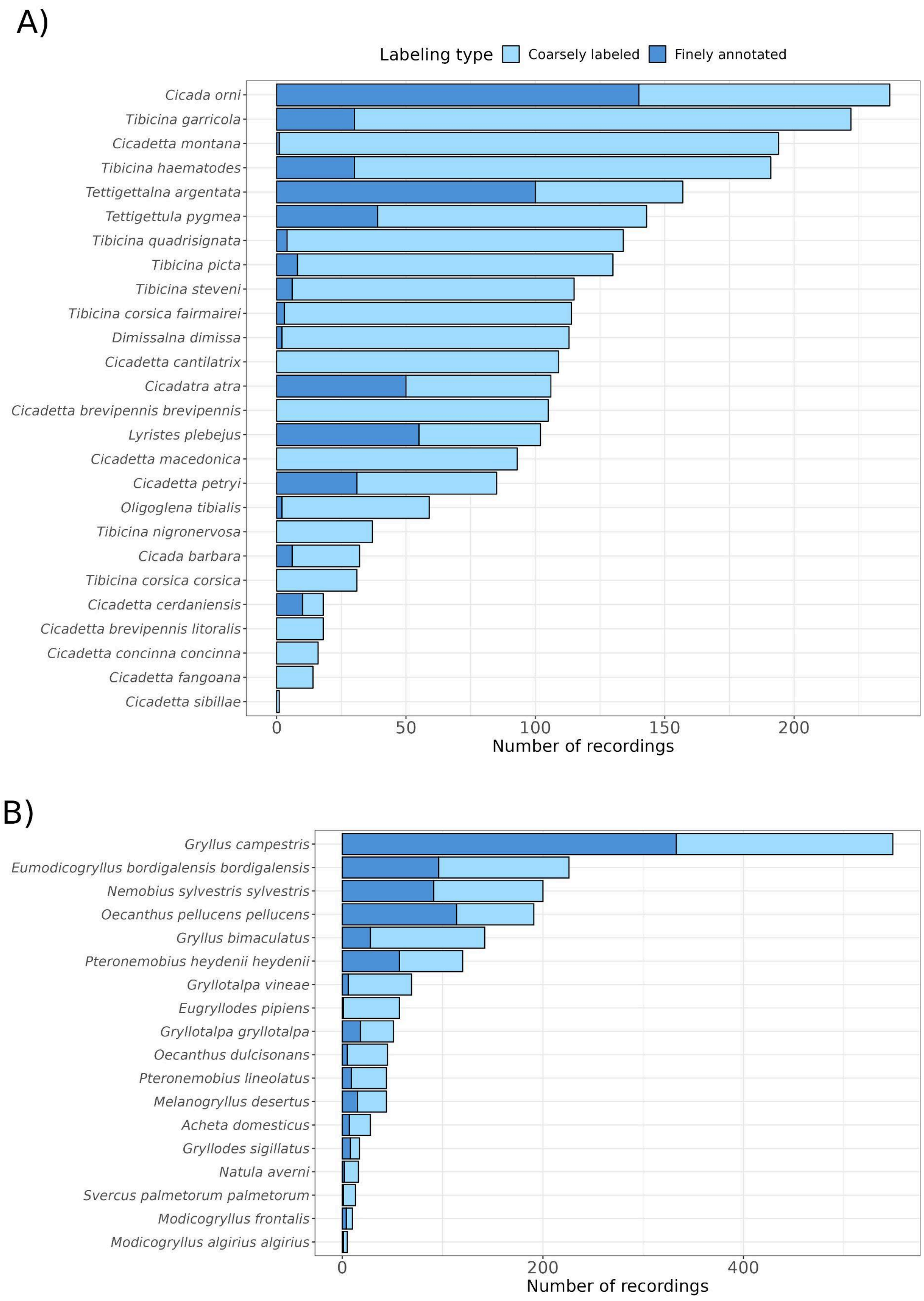

C)

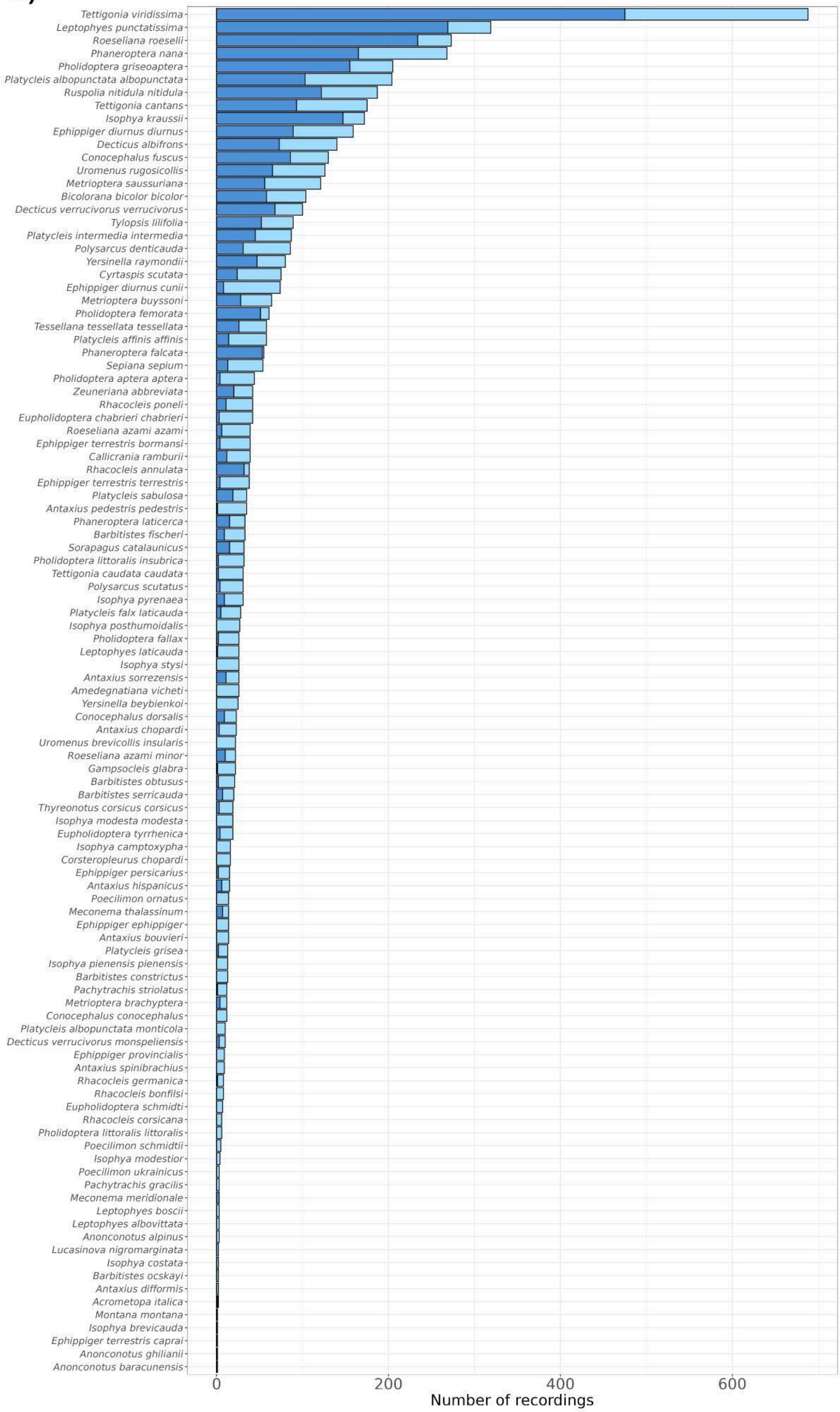

D)

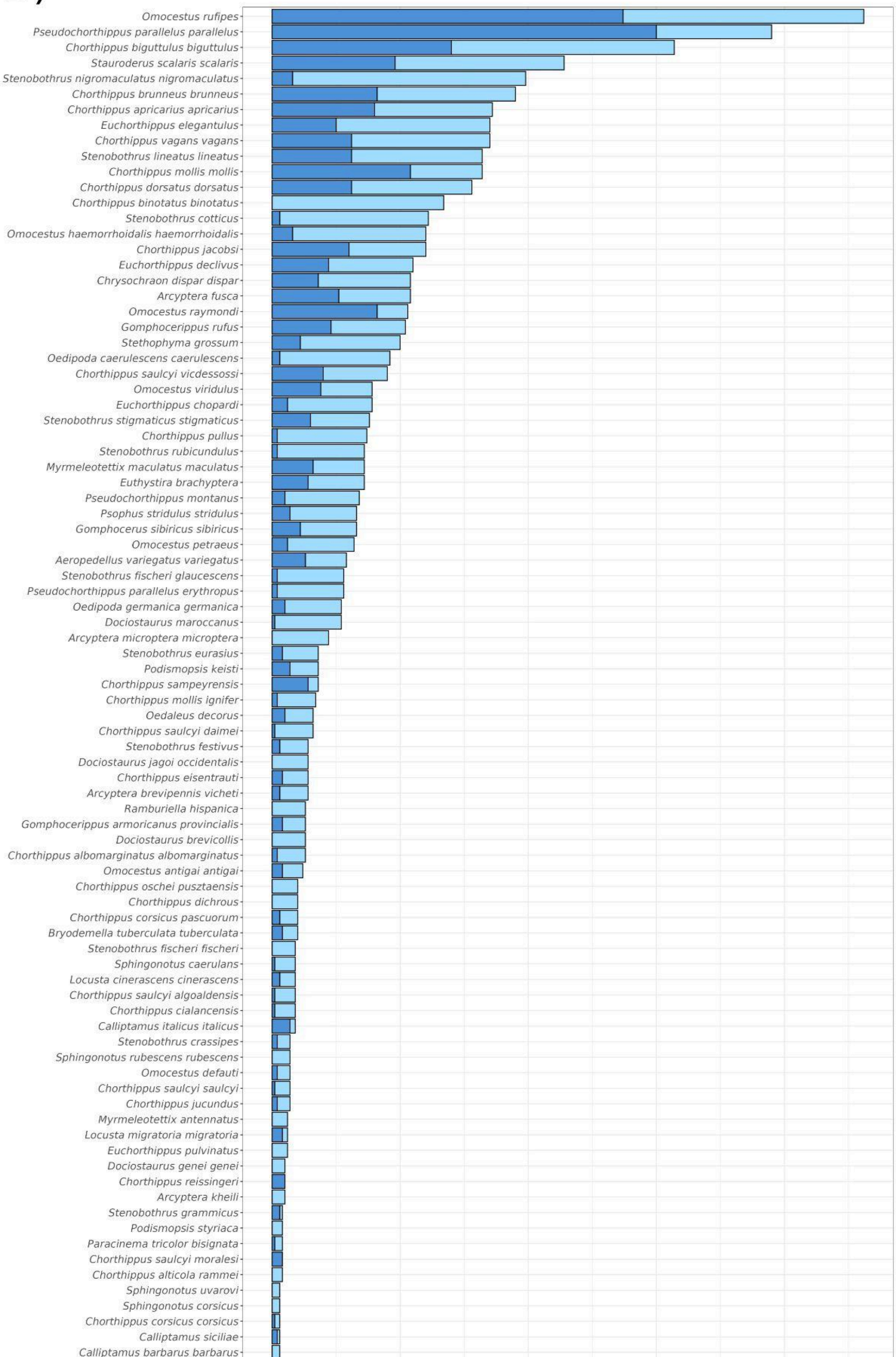

**Figure 3:** Total number of weakly (light blue) and strongly (dark blue) labeled recordings for each (A) cicada (Cicadidae), (B) cricket and mole cricket (Grylloidea and Gryllotalpoidea), (C) katydid (Tettigoniidae) and (D) grasshopper (Acridoidea) (sub)species in our dataset.

It is also important to note that the train/validation/test split proposed in this publication is intended as a suggestion. The Zenodo repository provides the complete set of original recordings, allowing other research teams to customize the dataset composition and subdivision according to their needs. This includes adjusting the train/validation/test split ratio, selecting a subset of locally occurring species or imposing an upper limit on annotations per species to further mitigate cross-species data imbalance.

## 5. Conclusion

Overall, we posit that the fine level of annotation provided for a third of our recordings, in combination with the aforementioned distinctive features of our dataset, could make it a valuable resource for the training of DL algorithms for the acoustic classification of orthopterans and cicadas in North, Central and temperate Western Europe. In addition, our dataset can also support other applications, such as extracting acoustic traits from the different sounds emitted by each species or analyzing regional variations in acoustic ecotypes. Future expansions will be added to the Zenodo repository, and we welcome contributions from entomologists, bioacousticians and ecoacousticians interested in enriching the dataset with additional recordings.

## CRediT authorship contribution statement

**David Funosas**: Writing – original draft, Visualization, Software, Methodology, Investigation, Formal analysis, Data curation. **Elodie Massol**: Writing – Review & Editing, Methodology, Investigation, Data curation. **Yves Bas**: Writing – Review & Editing, Investigation, Data curation. **Svenja Schmidt:** Writing – Review & Editing, Investigation, Data curation. **Dominik Arend**: Writing – Review & Editing, Investigation, Data curation. **Alexander Gebhard**: Writing – Review & Editing, Software, Validation. **Luc Barbaro**: Writing – Review & Editing. **Sebastian König**: Investigation, Data curation. **Rafael Carbonell Font**: Investigation, Data curation. **David Sannier**: Investigation, Data curation. **Fernand Deroussen**: Investigation, Data curation. **Jérôme Sueur**: Investigation, Data curation. **Christian Roesti**: Investigation, Data curation. **Tomi Trilar**: Investigation, Data curation.

**Wolfgang Forstmeier**: Investigation, Data curation. **Lucas Roger**: Investigation, Data curation. **Eloïsa Matheu**: Investigation, Data curation. **Piotr Guzik**: Investigation, Data curation. **Julien Barataud** : Investigation, Data curation. **Laurent Pelozuelo**: Investigation, Data curation. **Stéphane Puissant**: Investigation, Data curation. **Sandra Mueller**: Writing – Review & Editing. **Björn Schuller**: Writing – Review & Editing. **José Montoya**: Writing – Review & Editing. **Andreas Triantafyllopoulos**: Software. **Maxime Cauchoix**: Writing – Review & Editing, Validation, Supervision, Resources, Project Administration, Methodology, Funding Acquisition, Conceptualization.

## Acknowledgements

We are deeply grateful to Roy Kleukers, Mathieu Pélissié, Jakub Burdzicki, Margaux Charra, Adrien Charbonneau, Daniel Espejo Fraga, Joss Deffarges, Lukasz Cudziło, Daniel Bizet, Ghislain Riou, Marta Celej, Marie-Lilith Patou, Blandine Carre, Antoine Chabrolle, Joan Ventura Linares, Marc Corail, Matija Gogala, Mathieu Sannier, Vincent Milaret, Alexis Laforge, Pere Pons, Joan Estrada Bonell, Florence Matutini, Benjamin Drillat, Berenger Remy, Adeline Pichard, Evgenia Kovalyova, Laura Martin, Remi Jullian, Alexandre Crégu, Aurélie Torres, Christian Kerbiriou, Marlene Massouh, Nicolas Mokuenko, Jérome Allain, Romain Riols, Varvara Vedenina, Benoit Nabholz, Carlos Álvarez-Cros, Elouan Meyniel, Gaëtan Jouvenez, Georges Bedrines, Nicolas Vissyrias, Celine Quelennec, Clement Lemarchand, Clementine Azam, Eric Sardet, Klaus Alix, Rafael Tamajón, Sylvain Grimaud, Julien Cavallo, Leslie Campourcy, Sébastien Merle, Tamás Kiss, Xavier Béjar, Aurélien Grimaud, Fabien Sane, Jocelyn Fonderflick, Justine Przybilski, Marc Anton, Thomas Armand and Werner Reitmeier, who granted us permission to incorporate their recordings into our dataset (Table S2). We also want to thank Johanna Berger, Maren Teschauer, Benjamin Schmid, Orian Ly, Lutèce Mezzetta, Mathilde Lladó, Eva Blot and Léa Geng for their collaboration in the annotation of recordings, Alexander Teschke and Markus Rubenbauer for their help with maintaining recording sites, Thierry Feuillet for facilitating the collection of mountain orthopteran recordings through the SpatialTreeP project, and the www.ornitho.cat and sonotheque.mnhn.fr (from Muséum National d’Histoire Naturelle) platforms for facilitating the access to some of the recordings in our dataset.

## Funding

We declare having received funding from the Psi-Biom project (French PIA 3 under grant number 2182D0406-A), from the French National Program (ANR) “Investment for Future-Excellency Equipment” (project TERRA FORMA, with the reference ANR-21-ESRE-0014), from LabEx Tulip and from coauthor MC's discretionary funding from his Junior Professor Chair position Neo Sensation (ANR-23-CPJ1-0174-01). Acoustic research by coauthor TT was conducted as part of the programme “Communities, relationships, and communication in ecosystems” (No. P1-0255), funded by the Slovenian Research and Innovation Agency, and part of the acoustic research by coauthors EM and DF was conducted as part of the SpatialTreeP project funded by French ANR (ANR-21-CE03-0002).

## Conflict of interest disclosure

The authors declare that they comply with the PCI rule of having no financial conflicts of interest in relation to the content of the article.

## Data, script, code, and supplementary information availability

All data has been deposited in a Zenodo repository (https://doi.org/10.5281/zenodo.15043893) containing the following files:

· *recording_metadata.csv*, containing the metadata and corresponding license of each recording. The fields included in the CSV are the following:

- *recording_id*: numerical code identifying each recording
- *recording_file_name*: name of the corresponding audio file in *whole_recordings.zip*
- *author_name*: name of the author of the recording
- *recording_date*: date when the recording took place, in YYYY-mm-dd format
- *recording_time*: time of the day when the recording took place, in 24-hour format
- *recording_diel_period*: “day” if the recording took place between dawn and dusk, “night” otherwise
- *country_code*: country where the recording took place, in Alpha-2 code format
- *region*: region, state or department where the recording took place

- *commune*: commune or municipality where the recording took place
- *latitude*: latitude coordinate in WGS 84
- *longitude*: longitude coordinate in WGS 84
- *weather*: weather (“Sunny” or “Cloudy” for daytime recordings and “Cloudless” or “Cloudy” for nighttime ones) at the time and place of the recording
- *air_temperature*: air temperature at the time and place of the recording, in degrees Celsius
- *support_temperature*: temperature of the support from which the insect was stridulating or timbalizing, in degrees Celsius
- *recorder*: recording device used to record the sound
- *microphone*: microphone used to record the sound
- *duration_min*: truncated duration of the recording in minutes
- *duration_sec*: seconds to add to *duration_min* in order to get the full duration of the recording
- *sampling_rate*: sampling rate of the recording, in Hz
- *BPS*: bit rate of the recording measured in bits per second
- *audio_channels*: “mono” if the recording only has one audio channel, “stereo” if it has two channels
- *annotated*: “Exhaustively annotated” if all sounds in the recording have been annotated, “Partially annotated” if only a subset of sounds has been annotated and “Unannotated” if no sounds have been annotated
- *license*: license under which the recording and corresponding annotations can be used
- *recorded_species*: non-exhaustive list of orthopteran and cicada species present in the recording

· *online_recordings_metadata.csv*, equivalent to *recording_metadata.csv* but concerning orthopteran and cicada recordings available on the online libraries Xeno-canto, iNaturalist, observation.org, ZFMK, MinIO and BioAcoustica.

· *annotated_audio_segments.csv*, containing the time and frequency bounds of each annotation included in the dataset, as well as the set —train, validation or test— to which each audio segment has been assigned. The fields included in the CSV are the following:

- *recording_id*: numerical code identifying each recording
- *audio_segment_initial_time*: initial time, in seconds, of the audio segment relative to the beginning of the recording
- *audio_segment_final_time*: final time, in seconds, of the audio segment relative to the beginning of the recording

- *annotation_initial_time*: initial time, in seconds, of the original annotation relative to the beginning of the recording
- *annotation_final_time*: final time, in seconds, of the original annotation relative to the beginning of the recording
- *annotation_min_freq*: minimum frequency, in Hz, of the original annotation
- *annotation_max_freq*: maximum frequency, in Hz, of the original annotation
- *label*: label of the annotation, corresponding either to a scientific name for biotic sounds or to an ad hoc sound category for geological and anthropogenic sounds (Table S3)
- *label_category*: category corresponding to the taxonomic order of the species labeled for biotic sounds, to “Anthropophony” for anthropogenic sounds and to “Geophony” for abiotic sounds of natural origin such as wind or rain
- *subset*: subset (“train”, “val” or “test”) to which the audio segment has been assigned for the development of DL algorithms
- *audio_segment_file_name*: name of the audio file corresponding to the audio segment in *split_annotated_recordings.zip*

· *annotated_audio_segments_by_label_summary.csv*, containing a numeric overview of the train/validation/test division of audio segments for each label. The fields included in the CSV are the following:

- *label*: annotation label, corresponding either to a scientific name for biotic sounds or to an ad hoc sound category for geological and anthropogenic sounds (Table S3)
- *label_category*: equivalent to its analog field in *annotated_audio_segments.csv*
- *n_audio_segments_in_train*: number of audio segments in the train set where the label is present
- *n_audio_segments_in_val*: number of audio segments in the validation set where the label is present
- *n_audio_segments_in_test*: number of audio segments in the test set where the label is present

· *whole_recordings.zip*, containing all original recordings comprised in the dataset in WAV format. All files are found inside a folder named after the license they are made available under.

· *split_annotated_recordings.zip*, containing all the annotated 4-second audio segments comprised in the dataset in WAV format. All audio segments are found inside a folder named after the set (train/val/test) they have been assigned to.

Due to the variety of sources from which our recordings were drawn, different audio files and annotations are made available under three different licenses: the Creative Commons “Attribution” data waiver (CC BY 4.0), the Creative Commons “Attribution-Noncommercial” data waiver (CC BY-NC 4.0) and the Creative Commons “Attribution-NonCommercial-NoDerivatives” data waiver (CC BY-NC-ND 4.0). Additionally, some recordings are released without any license attached at the explicit request of their authors. These recordings are stored in an encrypted ZIP file, and potential users must contact the corresponding author to request access. The license —or lack thereof— assigned to each recording is described in the *recording_metadata.csv* file.

# Supplementary Material

| Species | Order | Sound type | Frequency range | Number of recordings on online libraries | Number of recordings in our dataset | Number of recordings annotated | Number of audio segments annotated in the train set | Number of audio segments annotated in the val set | Number of audio segments annotated in the test set |
|---|---|---|---|---|---|---|---|---|---|
| ***Cicada barbara sspp.*** | Hemiptera | Timbalization | Audible | 386 | 32 | 6 | 31 | 0 | 4 |
| ***Cicada orni*** | Hemiptera | Timbalization | Audible | 1203 | 237 | 140 | 549 | 172 | 228 |
| ***Cicadatra atra*** | Hemiptera | Timbalization | Audible | 47 | 106 | 50 | 192 | 77 | 48 |
| ***Cicadetta brevipennis brevipennis*** | Hemiptera | Timbalization | Partially audible | 12 | 105 | 0 | 0 | 0 | 0 |
| ***Cicadetta brevipennis litoralis*** | Hemiptera | Timbalization | Partially audible | 2 | 18 | 0 | 0 | 0 | 0 |
| ***Cicadetta cantilatrix*** | Hemiptera | Timbalization | Partially audible | 17 | 109 | 0 | 0 | 0 | 0 |
| ***Cicadetta cerdaniensis*** | Hemiptera | Timbalization | Partially audible | 0 | 18 | 10 | 82 | 0 | 17 |
| ***Cicadetta concinna concinna*** | Hemiptera | Timbalization | Partially audible | 0 | 16 | 0 | 0 | 0 | 0 |
| ***Cicadetta fangoana*** | Hemiptera | Timbalization | Partially audible | 2 | 14 | 0 | 0 | 0 | 0 |
| ***Cicadetta macedonica*** | Hemiptera | Timbalization | Partially audible | 5 | 93 | 0 | 0 | 0 | 0 |
| ***Cicadetta montana*** | Hemiptera | Timbalization | Partially audible | 11 | 194 | 1 | 9 | 0 | 0 |
| ***Cicadetta petryi*** | Hemiptera | Timbalization | Partially audible | 7 | 85 | 31 | 102 | 62 | 41 |
| ***Cicadetta sibillae*** | Hemiptera | Timbalization | Partially audible | 1 | 1 | 0 | 0 | 0 | 0 |
| ***Dimissalna dimissa*** | Hemiptera | Timbalization | Partially audible | 19 | 113 | 2 | 24 | 0 | 0 |
| ***Lyristes plebejus*** | Hemiptera | Timbalization | Audible | 114 | 102 | 55 | 285 | 58 | 106 |
| ***Oligoglena tibialis*** | Hemiptera | Timbalization | Partially audible | 14 | 59 | 2 | 3 | 0 | 0 |
| ***Tettigettalna argentata*** | Hemiptera | Timbalization | Audible | 282 | 157 | 100 | 411 | 121 | 99 |
| ***Tettigettula pygmea*** | Hemiptera | Timbalization | Partially audible | 37 | 143 | 39 | 175 | 61 | 40 |
| ***Tibicina corsica corsica*** | Hemiptera | Timbalization | Audible | 6 | 31 | 0 | 0 | 0 | 0 |
| ***Tibicina corsica fairmairei*** | Hemiptera | Timbalization | Audible | 1 | 114 | 3 | 60 | 0 | 7 |
| ***Tibicina garricola*** | Hemiptera | Timbalization | Audible | 63 | 222 | 30 | 175 | 95 | 44 |
| ***Tibicina haematodes*** | Hemiptera | Timbalization | Audible | 61 | 191 | 30 | 104 | 64 | 32 |
| ***Tibicina nigronervosa*** | Hemiptera | Timbalization | Audible | 0 | 37 | 0 | 0 | 0 | 0 |
| ***Tibicina picta*** | Hemiptera | Timbalization | Audible | 0 | 130 | 8 | 51 | 10 | 4 |
| ***Tibicina quadrisignata*** | Hemiptera | Timbalization | Audible | 1 | 134 | 4 | 45 | 0 | 23 |
| ***Tibicina steveni*** | Hemiptera | Timbalization | Audible | 18 | 115 | 6 | 10 | 2 | 42 |
| ***Acheta domesticus*** | Orthoptera | Stridulation | Audible | 300 | 28 | 7 | 49 | 0 | 10 |
| ***Acrometopa italica*** | Orthoptera | Stridulation | Audible | 12 | 2 | 1 | 11 | 0 | 0 |
| ***Acrotylus braudi*** | Orthoptera | Wing rattle | Audible | 0 | 0 | 0 | 0 | 0 | 0 |
| ***Acrotylus fischeri*** | Orthoptera | Wing rattle | Audible | 0 | 1 | 0 | 0 | 0 | 0 |
| ***Acrotylus insubricus braudi*** | Orthoptera | Wing rattle | Audible | 0 | 0 | 0 | 0 | 0 | 0 |
| ***Acrotylus insubricus insubricus*** | Orthoptera | Wing rattle | Audible | 0 | 1 | 0 | 0 | 0 | 0 |
| ***Acrotylus patruelis*** | Orthoptera | Wing rattle | Audible | 0 | 0 | 0 | 0 | 0 | 0 |
| ***Aeropedellus variegatus variegatus*** | Orthoptera | Stridulation | Audible | 18 | 29 | 13 | 18 | 0 | 5 |
| ***Aiolopus puissanti*** | Orthoptera | Wing rattle | Audible | 0 | 0 | 0 | 0 | 0 | 0 |
| ***Aiolopus strepens*** | Orthoptera | Wing rattle | Audible | 6 | 0 | 0 | 0 | 0 | 0 |
| ***Aiolopus thalassinus corsicus*** | Orthoptera | Wing rattle | Audible | 0 | 0 | 0 | 0 | 0 | 0 |
| ***Aiolopus thalassinus thalassinus*** | Orthoptera | Wing rattle | Audible | 0 | 0 | 0 | 0 | 0 | 0 |
| ***Amedegnatiana vicheti*** | Orthoptera | Stridulation | Ultrasounds | 8 | 26 | 0 | 0 | 0 | 0 |
| ***Anacridium aegyptium*** | Orthoptera | Wing rattle | Audible | 0 | 0 | 0 | 0 | 0 | 0 |
| ***Anonconotus alpinus*** | Orthoptera | Stridulation | Ultrasounds | 20 | 3 | 0 | 0 | 0 | 0 |
| ***Anonconotus baracunensis*** | Orthoptera | Stridulation | Ultrasounds | 12 | 1 | 0 | 0 | 0 | 0 |
| ***Anonconotus ghilianii*** | Orthoptera | Stridulation | Partially audible | 11 | 1 | 0 | 0 | 0 | 0 |
| ***Anonconotus italoaustriacus*** | Orthoptera | Stridulation | Ultrasounds | 11 | 0 | 0 | 0 | 0 | 0 |
| ***Anonconotus ligustinus*** | Orthoptera | Stridulation | Ultrasounds | 10 | 0 | 0 | 0 | 0 | 0 |
| ***Anonconotus mercantouri*** | Orthoptera | Stridulation | Ultrasounds | 10 | 0 | 0 | 0 | 0 | 0 |
| ***Anonconotus occidentalis*** | Orthoptera | Stridulation | Ultrasounds | 1 | 0 | 0 | 0 | 0 | 0 |
| ***Antaxius bouvieri*** | Orthoptera | Stridulation | Ultrasounds | 7 | 14 | 0 | 0 | 0 | 0 |
| ***Antaxius chopardi*** | Orthoptera | Stridulation | Ultrasounds | 20 | 23 | 3 | 14 | 0 | 4 |
| ***Antaxius difformis*** | Orthoptera | Stridulation | Ultrasounds | 14 | 2 | 0 | 0 | 0 | 0 |
| ***Antaxius hispanicus*** | Orthoptera | Stridulation | Ultrasounds | 38 | 15 | 6 | 13 | 9 | 6 |
| ***Antaxius pedestris pedestris*** | Orthoptera | Stridulation | Ultrasounds | 35 | 35 | 1 | 1 | 0 | 0 |
| ***Antaxius sorrezensis*** | Orthoptera | Stridulation | Ultrasounds | 4 | 26 | 11 | 7 | 2 | 2 |
| ***Antaxius spinibrachius*** | Orthoptera | Stridulation | Ultrasounds | 75 | 9 | 0 | 0 | 0 | 0 |
| ***Arcyptera alzonai*** | Orthoptera | Stridulation | Audible | 0 | 0 | 0 | 0 | 0 | 0 |
| ***Arcyptera brevipennis vicheti*** | Orthoptera | Stridulation | Audible | 13 | 14 | 3 | 3 | 0 | 0 |
| ***Arcyptera fusca*** | Orthoptera | Stridulation | Audible | 54 | 54 | 26 | 41 | 21 | 9 |
| ***Arcyptera kheili*** | Orthoptera | Stridulation | Audible | 11 | 5 | 0 | 0 | 0 | 0 |
| ***Arcyptera microptera carpentieri*** | Orthoptera | Stridulation | Audible | 9 | 3 | 0 | 0 | 0 | 0 |
| ***Arcyptera microptera microptera*** | Orthoptera | Stridulation | Audible | 8 | 22 | 0 | 0 | 0 | 0 |
| ***Barbitistes constrictus*** | Orthoptera | Stridulation | Partially audible | 35 | 13 | 0 | 0 | 0 | 0 |
| ***Barbitistes fischeri*** | Orthoptera | Stridulation | Partially audible | 48 | 33 | 9 | 29 | 22 | 8 |
| ***Barbitistes obtusus*** | Orthoptera | Stridulation | Partially audible | 41 | 21 | 2 | 13 | 0 | 4 |
| ***Barbitistes ocskayi*** | Orthoptera | Stridulation | Partially audible | 225 | 2 | 0 | 0 | 0 | 0 |
| ***Barbitistes serricauda*** | Orthoptera | Stridulation | Ultrasounds | 95 | 20 | 7 | 13 | 3 | 5 |
| ***Bicolorana bicolor bicolor*** | Orthoptera | Stridulation | Audible | 185 | 104 | 58 | 328 | 68 | 68 |
| ***Bryodemella tuberculata tuberculata*** | Orthoptera | Wing rattle | Audible | 1 | 10 | 4 | 17 | 0 | 8 |
| ***Callicrania ramburii*** | Orthoptera | Stridulation | Partially audible | 25 | 39 | 12 | 21 | 8 | 3 |
| ***Calliptamus barbarus barbarus*** | Orthoptera | Mandibular sounds | Audible | 0 | 3 | 0 | 0 | 0 | 0 |
| ***Calliptamus italicus italicus*** | Orthoptera | Mandibular sounds | Audible | 0 | 9 | 7 | 18 | 0 | 0 |
| ***Calliptamus siciliae*** | Orthoptera | Mandibular sounds | Audible | 0 | 3 | 2 | 10 | 0 | 0 |
| ***Calliptamus wattenwylianus*** | Orthoptera | Mandibular sounds | Audible | 0 | 0 | 0 | 0 | 0 | 0 |
| ***Chorthippus albomarginatus albomarginatus*** | Orthoptera | Stridulation | Audible | 78 | 13 | 2 | 4 | 0 | 0 |
| ***Chorthippus alticola rammei*** | Orthoptera | Stridulation | Audible | 0 | 4 | 0 | 0 | 0 | 0 |
| ***Chorthippus apricarius apricarius*** | Orthoptera | Stridulation | Audible | 150 | 86 | 40 | 156 | 25 | 28 |
| ***Chorthippus biguttulus biguttulus*** | Orthoptera | Stridulation | Audible | 725 | 157 | 70 | 199 | 96 | 122 |
| ***Chorthippus binotatus binotatus*** | Orthoptera | Stridulation | Audible | 14 | 67 | 0 | 0 | 0 | 0 |
| ***Chorthippus brunneus brunneus*** | Orthoptera | Stridulation | Audible | 390 | 95 | 41 | 66 | 19 | 25 |
| ***Chorthippus cialancensis*** | Orthoptera | Stridulation | Audible | 1 | 9 | 1 | 1 | 0 | 0 |
| ***Chorthippus corsicus corsicus*** | Orthoptera | Stridulation | Audible | 5 | 3 | 1 | 11 | 0 | 0 |
| ***Chorthippus corsicus pascuorum*** | Orthoptera | Stridulation | Audible | 8 | 10 | 3 | 3 | 0 | 0 |
| ***Chorthippus dichrous*** | Orthoptera | Stridulation | Audible | 7 | 10 | 0 | 0 | 0 | 0 |
| ***Chorthippus dorsatus dorsatus*** | Orthoptera | Stridulation | Audible | 118 | 78 | 31 | 74 | 34 | 20 |
| ***Chorthippus eisentrauti*** | Orthoptera | Stridulation | Audible | 182 | 14 | 4 | 6 | 0 | 2 |
| ***Chorthippus jacobsi*** | Orthoptera | Stridulation | Audible | 76 | 60 | 30 | 55 | 28 | 29 |
| ***Chorthippus jucundus*** | Orthoptera | Stridulation | Audible | 14 | 7 | 2 | 2 | 0 | 2 |
| ***Chorthippus jutlandica*** | Orthoptera | Stridulation | Audible | 2 | 0 | 0 | 0 | 0 | 0 |
| ***Chorthippus mollis ignifer*** | Orthoptera | Stridulation | Audible | 113 | 17 | 2 | 33 | 0 | 3 |
| ***Chorthippus mollis mollis*** | Orthoptera | Stridulation | Audible | 270 | 82 | 54 | 125 | 44 | 35 |
| ***Chorthippus oschei pusztaensis*** | Orthoptera | Stridulation | Audible | 0 | 10 | 0 | 0 | 0 | 0 |
| ***Chorthippus pullus*** | Orthoptera | Stridulation | Audible | 18 | 37 | 2 | 2 | 0 | 0 |
| ***Chorthippus reissingeri*** | Orthoptera | Stridulation | Audible | 11 | 5 | 5 | 9 | 0 | 15 |
| ***Chorthippus sampeyrensis*** | Orthoptera | Stridulation | Audible | 1 | 18 | 14 | 23 | 0 | 0 |
| ***Chorthippus saulcyi algoaldensis*** | Orthoptera | Stridulation | Audible | 3 | 9 | 1 | 1 | 0 | 0 |
| ***Chorthippus saulcyi daimei*** | Orthoptera | Stridulation | Audible | 3 | 16 | 1 | 1 | 0 | 0 |
| ***Chorthippus saulcyi moralesi*** | Orthoptera | Stridulation | Audible | 4 | 4 | 4 | 31 | 0 | 3 |
| ***Chorthippus saulcyi saulcyi*** | Orthoptera | Stridulation | Audible | 0 | 7 | 1 | 1 | 0 | 0 |
| ***Chorthippus saulcyi vicdessossi*** | Orthoptera | Stridulation | Audible | 2 | 45 | 20 | 72 | 18 | 17 |
| ***Chorthippus smardai*** | Orthoptera | Stridulation | Audible | 0 | 0 | 0 | 0 | 0 | 0 |
| ***Chorthippus vagans vagans*** | Orthoptera | Stridulation | Audible | 130 | 85 | 31 | 62 | 38 | 4 |
| ***Chrysochraon dispar dispar*** | Orthoptera | Stridulation | Audible | 154 | 54 | 18 | 10 | 42 | 12 |
| ***Conocephalus conocephalus*** | Orthoptera | Stridulation | Ultrasounds | 18 | 12 | 0 | 0 | 0 | 0 |
| ***Conocephalus dorsalis*** | Orthoptera | Stridulation | Partially audible | 85 | 23 | 9 | 29 | 7 | 10 |
| ***Conocephalus fuscus*** | Orthoptera | Stridulation | Partially audible | 419 | 130 | 86 | 530 | 262 | 102 |
| ***Corsteropleurus chopardi*** | Orthoptera | Stridulation | Audible | 18 | 16 | 0 | 0 | 0 | 0 |
| ***Cyrtaspis scutata*** | Orthoptera | Stridulation | Ultrasounds | 98 | 75 | 24 | 22 | 8 | 5 |
| ***Decticus albifrons*** | Orthoptera | Stridulation | Audible | 304 | 140 | 73 | 224 | 80 | 73 |
| ***Decticus verrucivorus monspeliensis*** | Orthoptera | Stridulation | Audible | 3 | 10 | 3 | 21 | 0 | 0 |
| ***Decticus verrucivorus verrucivorus*** | Orthoptera | Stridulation | Audible | 125 | 100 | 68 | 237 | 110 | 37 |
| ***Dociostaurus brevicollis*** | Orthoptera | Stridulation | Audible | 0 | 13 | 0 | 0 | 0 | 0 |
| ***Dociostaurus genei genei*** | Orthoptera | Stridulation | Audible | 1 | 5 | 0 | 0 | 0 | 0 |
| ***Dociostaurus jagoi occidentalis*** | Orthoptera | Stridulation | Audible | 5 | 14 | 0 | 0 | 0 | 0 |
| ***Dociostaurus maroccanus*** | Orthoptera | Stridulation | Audible | 20 | 27 | 1 | 7 | 0 | 0 |
| ***Ephippiger diurnus cunii*** | Orthoptera | Stridulation | Partially audible | 40 | 74 | 8 | 26 | 4 | 5 |
| ***Ephippiger diurnus diurnus*** | Orthoptera | Stridulation | Partially audible | 57 | 159 | 89 | 240 | 133 | 66 |
| ***Ephippiger ephippiger*** | Orthoptera | Stridulation | Partially audible | 26 | 14 | 0 | 0 | 0 | 0 |
| ***Ephippiger persicarius*** | Orthoptera | Stridulation | Partially audible | 2 | 15 | 2 | 15 | 0 | 0 |
| ***Ephippiger provincialis*** | Orthoptera | Stridulation | Partially audible | 12 | 9 | 0 | 0 | 0 | 0 |
| ***Ephippiger terrestris bormansi*** | Orthoptera | Stridulation | Partially audible | 12 | 39 | 4 | 29 | 0 | 0 |
| ***Ephippiger terrestris caprai*** | Orthoptera | Stridulation | Partially audible | 1 | 1 | 0 | 0 | 0 | 0 |
| ***Ephippiger terrestris terrestris*** | Orthoptera | Stridulation | Partially audible | 22 | 38 | 4 | 20 | 0 | 11 |
| ***Euchorthippus chopardi*** | Orthoptera | Stridulation | Audible | 36 | 39 | 6 | 8 | 0 | 0 |
| ***Euchorthippus declivus*** | Orthoptera | Stridulation | Audible | 64 | 55 | 22 | 41 | 12 | 15 |
| ***Euchorthippus elegantulus*** | Orthoptera | Stridulation | Audible | 29 | 85 | 25 | 54 | 23 | 19 |
| ***Euchorthippus pulvinatus*** | Orthoptera | Stridulation | Audible | 0 | 6 | 0 | 0 | 0 | 0 |
| ***Eugryllodes pipiens*** | Orthoptera | Stridulation | Audible | 100 | 57 | 1 | 4 | 0 | 0 |
| ***Eumodicogryllus bordigalensis bordigalensis*** | Orthoptera | Stridulation | Audible | 352 | 226 | 96 | 1512 | 295 | 241 |
| ***Eupholidoptera chabrieri chabrieri*** | Orthoptera | Stridulation | Partially audible | 37 | 42 | 3 | 9 | 0 | 0 |
| ***Eupholidoptera schmidti*** | Orthoptera | Stridulation | Partially audible | 34 | 7 | 0 | 0 | 0 | 0 |
| ***Eupholidoptera tyrrhenica*** | Orthoptera | Stridulation | Partially audible | 9 | 19 | 4 | 35 | 0 | 4 |
| ***Euthystira brachyptera*** | Orthoptera | Stridulation | Partially audible | 49 | 36 | 14 | 54 | 26 | 15 |
| ***Gampsocleis glabra*** | Orthoptera | Stridulation | Audible | 73 | 22 | 1 | 5 | 0 | 0 |
| ***Gomphoceridius brevipennis*** | Orthoptera | Stridulation | Audible | 0 | 2 | 0 | 0 | 0 | 0 |
| ***Gomphocerippus armoricanus provincialis*** | Orthoptera | Stridulation | Audible | 0 | 13 | 4 | 11 | 0 | 5 |
| ***Gomphocerippus rufus*** | Orthoptera | Stridulation | Audible | 104 | 52 | 23 | 14 | 6 | 55 |
| ***Gomphocerus sibiricus sibiricus*** | Orthoptera | Stridulation | Audible | 53 | 33 | 11 | 65 | 2 | 18 |
| ***Gryllodes sigillatus*** | Orthoptera | Stridulation | Audible | 63 | 17 | 8 | 107 | 0 | 12 |
| ***Gryllotalpa africana*** | Orthoptera | Stridulation | Audible | 46 | 0 | 0 | 0 | 0 | 0 |
| ***Gryllotalpa gryllotalpa*** | Orthoptera | Stridulation | Audible | 402 | 51 | 18 | 156 | 93 | 5 |
| ***Gryllotalpa vineae*** | Orthoptera | Stridulation | Audible | 139 | 69 | 6 | 16 | 5 | 7 |
| ***Gryllus bimaculatus*** | Orthoptera | Stridulation | Audible | 266 | 142 | 28 | 168 | 62 | 41 |
| ***Gryllus campestris*** | Orthoptera | Stridulation | Audible | 1057 | 549 | 333 | 2155 | 1104 | 917 |
| ***Isophya brevicauda*** | Orthoptera | Stridulation | Partially audible | 16 | 1 | 0 | 0 | 0 | 0 |
| ***Isophya camptoxypha*** | Orthoptera | Stridulation | Partially audible | 9 | 16 | 0 | 0 | 0 | 0 |
| ***Isophya costata*** | Orthoptera | Stridulation | Partially audible | 13 | 2 | 0 | 0 | 0 | 0 |
| ***Isophya kraussii*** | Orthoptera | Stridulation | Partially audible | 71 | 172 | 147 | 950 | 497 | 469 |
| ***Isophya modesta modesta*** | Orthoptera | Stridulation | Partially audible | 0 | 19 | 0 | 0 | 0 | 0 |
| ***Isophya modestior*** | Orthoptera | Stridulation | Partially audible | 99 | 4 | 0 | 0 | 0 | 0 |
| ***Isophya pienensis austromoravica*** | Orthoptera | Stridulation | Partially audible | 0 | 0 | 0 | 0 | 0 | 0 |
| ***Isophya pienensis pienensis*** | Orthoptera | Stridulation | Partially audible | 14 | 13 | 0 | 0 | 0 | 0 |
| ***Isophya pienensis sudetica*** | Orthoptera | Stridulation | Partially audible | 0 | 0 | 0 | 0 | 0 | 0 |
| ***Isophya posthumoidalis*** | Orthoptera | Stridulation | Partially audible | 11 | 27 | 0 | 0 | 0 | 0 |
| ***Isophya pyrenaea*** | Orthoptera | Stridulation | Partially audible | 37 | 31 | 9 | 12 | 1 | 3 |
| ***Isophya rectipennis*** | Orthoptera | Stridulation | Partially audible | 28 | 0 | 0 | 0 | 0 | 0 |
| ***Isophya stysi*** | Orthoptera | Stridulation | Partially audible | 6 | 26 | 0 | 0 | 0 | 0 |
| ***Leptophyes albovittata*** | Orthoptera | Stridulation | Ultrasounds | 34 | 3 | 0 | 0 | 0 | 0 |
| ***Leptophyes boscii*** | Orthoptera | Stridulation | Ultrasounds | 43 | 3 | 0 | 0 | 0 | 0 |
| ***Leptophyes laticauda*** | Orthoptera | Stridulation | Partially audible | 95 | 26 | 1 | 21 | 0 | 0 |
| ***Leptophyes punctatissima*** | Orthoptera | Stridulation | Ultrasounds | 1449 | 319 | 269 | 537 | 287 | 126 |
| ***Locusta cinerascens cinerascens*** | Orthoptera | Wing rattle | Audible | 0 | 9 | 3 | 2 | 0 | 1 |
| ***Locusta migratoria gallica*** | Orthoptera | Wing rattle | Audible | 0 | 0 | 0 | 0 | 0 | 0 |
| ***Locusta migratoria migratoria*** | Orthoptera | Wing rattle | Audible | 0 | 6 | 4 | 12 | 0 | 0 |
| ***Lucasinova nigromarginata*** | Orthoptera | Stridulation | Partially audible | 5 | 2 | 0 | 0 | 0 | 0 |
| ***Meconema meridionale*** | Orthoptera | Drumming | Audible | 14 | 3 | 3 | 26 | 0 | 3 |
| ***Meconema thalassinum*** | Orthoptera | Drumming | Audible | 8 | 14 | 7 | 10 | 0 | 1 |
| ***Melanogryllus desertus*** | Orthoptera | Stridulation | Audible | 54 | 44 | 15 | 22 | 8 | 14 |
| ***Metaplastes pulchripennis*** | Orthoptera | Stridulation | Ultrasounds | 25 | 0 | 0 | 0 | 0 | 0 |
| ***Metrioptera brachyptera*** | Orthoptera | Stridulation | Partially audible | 125 | 12 | 4 | 9 | 4 | 4 |
| ***Metrioptera buyssoni*** | Orthoptera | Stridulation | Partially audible | 25 | 64 | 28 | 84 | 35 | 41 |
| ***Metrioptera saussuriana*** | Orthoptera | Stridulation | Partially audible | 111 | 121 | 56 | 260 | 150 | 56 |
| ***Modicogryllus algirius algirius*** | Orthoptera | Stridulation | Audible | 10 | 5 | 1 | 5 | 0 | 0 |
| ***Modicogryllus frontalis*** | Orthoptera | Stridulation | Audible | 16 | 10 | 4 | 20 | 0 | 4 |
| ***Montana montana*** | Orthoptera | Stridulation | Audible | 0 | 1 | 0 | 0 | 0 | 0 |
| ***Myrmeleotettix antennatus*** | Orthoptera | Stridulation | Audible | 0 | 6 | 0 | 0 | 0 | 0 |
| ***Myrmeleotettix maculatus maculatus*** | Orthoptera | Stridulation | Audible | 106 | 36 | 16 | 111 | 0 | 0 |
| ***Natula averni*** | Orthoptera | Stridulation | Audible | 44 | 16 | 2 | 15 | 0 | 0 |
| ***Nemobius sylvestris sylvestris*** | Orthoptera | Stridulation | Audible | 276 | 200 | 91 | 414 | 129 | 130 |
| ***Oecanthus dulcisonans*** | Orthoptera | Stridulation | Audible | 186 | 45 | 5 | 40 | 22 | 15 |
| ***Oecanthus pellucens pellucens*** | Orthoptera | Stridulation | Audible | 763 | 191 | 114 | 834 | 584 | 261 |
| ***Oedaleus decorus*** | Orthoptera | Wing rattle | Audible | 0 | 16 | 5 | 19 | 0 | 0 |
| ***Oedipoda caerulescens caerulescens*** | Orthoptera | Wing rattle | Audible | 12 | 46 | 3 | 4 | 0 | 2 |
| ***Oedipoda caerulescens sardeti*** | Orthoptera | Wing rattle | Audible | 0 | 0 | 0 | 0 | 0 | 0 |
| ***Oedipoda charpentieri*** | Orthoptera | Wing rattle | Audible | 0 | 0 | 0 | 0 | 0 | 0 |
| ***Oedipoda coerulea*** | Orthoptera | Wing rattle | Audible | 0 | 2 | 2 | 12 | 0 | 1 |
| ***Oedipoda fuscocincta morini*** | Orthoptera | Wing rattle | Audible | 0 | 0 | 0 | 0 | 0 | 0 |
| ***Oedipoda germanica germanica*** | Orthoptera | Wing rattle | Audible | 4 | 27 | 5 | 11 | 0 | 5 |
| ***Omocestus antigai antigai*** | Orthoptera | Stridulation | Audible | 7 | 12 | 4 | 7 | 2 | 3 |
| ***Omocestus antigai bellmanni*** | Orthoptera | Stridulation | Audible | 7 | 1 | 1 | 10 | 0 | 0 |
| ***Omocestus defauti*** | Orthoptera | Stridulation | Audible | 8 | 7 | 2 | 0 | 0 | 5 |
| ***Omocestus haemorrhoidalis haemorrhoidalis*** | Orthoptera | Stridulation | Audible | 37 | 60 | 8 | 5 | 0 | 20 |
| ***Omocestus petraeus*** | Orthoptera | Stridulation | Audible | 25 | 32 | 6 | 12 | 0 | 4 |
| ***Omocestus raymondi*** | Orthoptera | Stridulation | Audible | 53 | 53 | 41 | 52 | 34 | 10 |
| ***Omocestus rufipes*** | Orthoptera | Stridulation | Audible | 110 | 231 | 137 | 235 | 70 | 63 |
| ***Omocestus viridulus*** | Orthoptera | Stridulation | Audible | 196 | 39 | 19 | 102 | 68 | 21 |
| ***Pachytrachis gracilis*** | Orthoptera | Stridulation | Audible | 42 | 3 | 0 | 0 | 0 | 0 |
| ***Pachytrachis striolatus*** | Orthoptera | Stridulation | Audible | 28 | 12 | 1 | 5 | 0 | 0 |
| ***Paracinema tricolor bisignata*** | Orthoptera | Wing rattle | Audible | 0 | 4 | 1 | 30 | 0 | 0 |
| ***Phaneroptera falcata*** | Orthoptera | Stridulation | Partially audible | 385 | 55 | 53 | 226 | 127 | 63 |
| ***Phaneroptera laticerca*** | Orthoptera | Stridulation | Partially audible | 395 | 33 | 15 | 62 | 20 | 8 |
| ***Phaneroptera nana*** | Orthoptera | Stridulation | Partially audible | 416 | 268 | 165 | 221 | 153 | 74 |
| ***Pholidoptera aptera aptera*** | Orthoptera | Stridulation | Partially audible | 71 | 44 | 4 | 28 | 0 | 0 |
| ***Pholidoptera fallax*** | Orthoptera | Stridulation | Partially audible | 40 | 26 | 2 | 4 | 0 | 0 |
| ***Pholidoptera femorata*** | Orthoptera | Stridulation | Partially audible | 75 | 61 | 51 | 176 | 58 | 41 |
| ***Pholidoptera griseoaptera*** | Orthoptera | Stridulation | Partially audible | 1649 | 205 | 155 | 709 | 338 | 261 |
| ***Pholidoptera littoralis insubrica*** | Orthoptera | Stridulation | Audible | 0 | 32 | 2 | 9 | 0 | 0 |
| ***Pholidoptera littoralis littoralis*** | Orthoptera | Stridulation | Audible | 35 | 6 | 0 | 0 | 0 | 0 |
| ***Platycleis affinis affinis*** | Orthoptera | Stridulation | Partially audible | 154 | 58 | 14 | 142 | 17 | 13 |
| ***Platycleis albopunctata albopunctata*** | Orthoptera | Stridulation | Partially audible | 138 | 204 | 103 | 330 | 94 | 109 |
| ***Platycleis albopunctata monticola*** | Orthoptera | Stridulation | Partially audible | 10 | 10 | 0 | 0 | 0 | 0 |
| ***Platycleis falx laticauda*** | Orthoptera | Stridulation | Partially audible | 11 | 28 | 5 | 95 | 0 | 1 |
| ***Platycleis grisea*** | Orthoptera | Stridulation | Partially audible | 68 | 13 | 2 | 4 | 0 | 0 |
| ***Platycleis intermedia intermedia*** | Orthoptera | Stridulation | Partially audible | 167 | 87 | 45 | 191 | 46 | 46 |
| ***Platycleis sabulosa*** | Orthoptera | Stridulation | Partially audible | 51 | 35 | 19 | 46 | 14 | 8 |
| ***Podismopsis keisti*** | Orthoptera | Stridulation | Audible | 2 | 18 | 7 | 37 | 0 | 6 |
| ***Podismopsis styriaca*** | Orthoptera | Stridulation | Audible | 0 | 4 | 0 | 0 | 0 | 0 |
| ***Poecilimon amissus*** | Orthoptera | Stridulation | Partially audible | 26 | 0 | 0 | 0 | 0 | 0 |
| ***Poecilimon gracilis*** | Orthoptera | Stridulation | Partially audible | 10 | 0 | 0 | 0 | 0 | 0 |
| ***Poecilimon ornatus*** | Orthoptera | Stridulation | Partially audible | 92 | 14 | 0 | 0 | 0 | 0 |
| ***Poecilimon schmidtii*** | Orthoptera | Stridulation | Partially audible | 12 | 5 | 0 | 0 | 0 | 0 |
| ***Poecilimon ukrainicus*** | Orthoptera | Stridulation | Partially audible | 1 | 3 | 0 | 0 | 0 | 0 |
| ***Polysarcus denticauda*** | Orthoptera | Stridulation | Partially audible | 100 | 86 | 31 | 151 | 116 | 7 |
| ***Polysarcus scutatus*** | Orthoptera | Stridulation | Partially audible | 29 | 31 | 4 | 18 | 4 | 7 |
| ***Pseudochorthippus montanus*** | Orthoptera | Stridulation | Audible | 66 | 34 | 5 | 32 | 0 | 10 |
| ***Pseudochorthippus parallelus erythropus*** | Orthoptera | Stridulation | Audible | 9 | 28 | 2 | 16 | 0 | 0 |
| ***Pseudochorthippus parallelus parallelus*** | Orthoptera | Stridulation | Audible | 334 | 195 | 150 | 600 | 220 | 226 |
| ***Psophus stridulus stridulus*** | Orthoptera | Wing rattle | Audible | 23 | 33 | 7 | 5 | 0 | 11 |
| ***Pteronemobius heydenii heydenii*** | Orthoptera | Stridulation | Audible | 105 | 120 | 57 | 390 | 159 | 84 |
| ***Pteronemobius lineolatus*** | Orthoptera | Stridulation | Audible | 33 | 44 | 9 | 132 | 0 | 19 |
| ***Ramburiella hispanica*** | Orthoptera | Stridulation | Audible | 20 | 13 | 0 | 0 | 0 | 0 |
| ***Rhacocleis annulata*** | Orthoptera | Stridulation | Partially audible | 29 | 38 | 32 | 67 | 46 | 15 |
| ***Rhacocleis bonfilsi*** | Orthoptera | Stridulation | Ultrasounds | 8 | 8 | 0 | 0 | 0 | 0 |
| ***Rhacocleis corsicana*** | Orthoptera | Stridulation | Ultrasounds | 29 | 6 | 0 | 0 | 0 | 0 |
| ***Rhacocleis germanica*** | Orthoptera | Stridulation | Ultrasounds | 105 | 8 | 1 | 2 | 0 | 0 |
| ***Rhacocleis poneli*** | Orthoptera | Stridulation | Partially audible | 20 | 42 | 11 | 6 | 2 | 3 |
| ***Roeseliana azami azami*** | Orthoptera | Stridulation | Partially audible | 11 | 39 | 6 | 59 | 0 | 4 |
| ***Roeseliana azami minor*** | Orthoptera | Stridulation | Partially audible | 7 | 22 | 10 | 42 | 7 | 5 |
| ***Roeseliana roeselii*** | Orthoptera | Stridulation | Partially audible | 543 | 273 | 234 | 1000 | 416 | 263 |
| ***Ruspolia nitidula nitidula*** | Orthoptera | Stridulation | Partially audible | 283 | 187 | 122 | 469 | 265 | 99 |
| ***Sepiana sepium*** | Orthoptera | Stridulation | Partially audible | 122 | 54 | 13 | 20 | 13 | 9 |
| ***Sorapagus catalaunicus*** | Orthoptera | Stridulation | Partially audible | 62 | 32 | 15 | 58 | 10 | 13 |
| ***Sphingonotus caerulans*** | Orthoptera | Stridulation | Audible | 12 | 9 | 1 | 6 | 0 | 0 |
| ***Sphingonotus corsicus*** | Orthoptera | Stridulation | Audible | 5 | 3 | 0 | 0 | 0 | 0 |
| ***Sphingonotus rubescens rubescens*** | Orthoptera | Stridulation | Audible | 6 | 7 | 0 | 0 | 0 | 0 |
| ***Sphingonotus uvarovi*** | Orthoptera | Stridulation | Audible | 0 | 3 | 0 | 0 | 0 | 0 |
| ***Stauroderus scalaris scalaris*** | Orthoptera | Stridulation | Audible | 114 | 114 | 48 | 84 | 47 | 21 |
| ***Stenobothrus cotticus*** | Orthoptera | Stridulation | Audible | 32 | 61 | 3 | 5 | 0 | 3 |
| ***Stenobothrus crassipes*** | Orthoptera | Stridulation | Audible | 0 | 7 | 2 | 9 | 0 | 0 |
| ***Stenobothrus eurasius*** | Orthoptera | Stridulation | Audible | 3 | 18 | 4 | 20 | 0 | 0 |
| ***Stenobothrus festivus*** | Orthoptera | Stridulation | Audible | 28 | 14 | 3 | 8 | 0 | 0 |
| ***Stenobothrus fischeri fischeri*** | Orthoptera | Stridulation | Audible | 12 | 9 | 0 | 0 | 0 | 0 |
| ***Stenobothrus fischeri glaucescens*** | Orthoptera | Stridulation | Audible | 4 | 28 | 2 | 13 | 0 | 0 |
| ***Stenobothrus grammicus*** | Orthoptera | Stridulation | Audible | 8 | 4 | 3 | 9 | 0 | 0 |
| ***Stenobothrus lineatus lineatus*** | Orthoptera | Stridulation | Audible | 123 | 82 | 31 | 158 | 46 | 56 |
| ***Stenobothrus nigromaculatus nigromaculatus*** | Orthoptera | Stridulation | Audible | 49 | 99 | 8 | 26 | 0 | 4 |
| ***Stenobothrus rubicundulus*** | Orthoptera | Stridulation | Audible | 53 | 36 | 2 | 8 | 0 | 0 |
| ***Stenobothrus stigmaticus stigmaticus*** | Orthoptera | Stridulation | Audible | 76 | 38 | 15 | 51 | 17 | 33 |
| ***Stethophyma grossum*** | Orthoptera | Stridulation | Audible | 78 | 50 | 11 | 19 | 3 | 3 |
| ***Svercus palmetorum palmetorum*** | Orthoptera | Stridulation | Audible | 96 | 13 | 1 | 3 | 0 | 0 |
| ***Tessellana tessellata tessellata*** | Orthoptera | Stridulation | Ultrasounds | 110 | 58 | 26 | 31 | 9 | 9 |
| ***Tessellana veyseli*** | Orthoptera | Stridulation | Ultrasounds | 3 | 0 | 0 | 0 | 0 | 0 |
| ***Tettigonia cantans*** | Orthoptera | Stridulation | Audible | 539 | 175 | 93 | 534 | 234 | 107 |
| ***Tettigonia caudata caudata*** | Orthoptera | Stridulation | Audible | 53 | 31 | 2 | 3 | 0 | 3 |
| ***Tettigonia viridissima*** | Orthoptera | Stridulation | Audible | 1527 | 688 | 475 | 2211 | 801 | 647 |
| ***Thyreonotus corsicus corsicus*** | Orthoptera | Stridulation | Partially audible | 47 | 19 | 3 | 9 | 0 | 3 |
| ***Tropidopola cylindrica cylindrica*** | Orthoptera | Wing rattle | Audible | 0 | 1 | 1 | 5 | 0 | 0 |
| ***Tylopsis lilifolia*** | Orthoptera | Stridulation | Partially audible | 104 | 89 | 52 | 74 | 40 | 22 |
| ***Uromenus brevicollis insularis*** | Orthoptera | Stridulation | Partially audible | 35 | 22 | 0 | 0 | 0 | 0 |
| ***Uromenus rugosicollis*** | Orthoptera | Stridulation | Audible | 150 | 126 | 65 | 207 | 83 | 40 |
| ***Yersinella beybienkoi*** | Orthoptera | Stridulation | Ultrasounds | 24 | 25 | 0 | 0 | 0 | 0 |
| ***Yersinella raymondii*** | Orthoptera | Stridulation | Ultrasounds | 182 | 80 | 47 | 46 | 21 | 19 |
| ***Zeuneriana abbreviata*** | Orthoptera | Stridulation | Partially audible | 72 | 42 | 20 | 236 | 51 | 16 |

**Table S1:** Type of sound produced by each known soniferous orthopteran and cicada species or subspecies in North, Central and temperate Western Europe (Andorra, Austria, Belgium, Czechia, Denmark, Estonia, Finland, mainland France and Corsica, Germany, Ireland, Latvia, Lithuania, Luxembourg, Monaco, Netherlands, Norway, Poland, United Kingdom, Sweden and Switzerland), along with a numerical overview of the data available both on online libraries (Xeno-canto, iNaturalist, observation.org, ZFMK, MinIO and BioAcoustica) and in our acoustic dataset for each species. Recordings identified only at the species level are excluded from subspecies counts, and heterodyne recordings, recordings without any license attached

and recordings whose identifications have not yet been validated are excluded from online recording counts.

| Recordist | Number of cicada recordings | Number of cicada species | Number of orthopteran recordings | Number of orthopteran species |
|---|---|---|---|---|
| **Adeline Pichard** | 0 | 0 | 8 | 2 |
| **Adrien Charbonneau** | 41 | 14 | 1 | 1 |
| **Alexandre Crégu** | 0 | 0 | 7 | 5 |
| **Alexis Laforge** | 0 | 0 | 13 | 3 |
| **Anonymous** | 0 | 0 | 1 | 1 |
| **Antoine Chabrolle** | 0 | 0 | 27 | 4 |
| **Aurélie Torres** | 3 | 1 | 4 | 4 |
| **Aurélien Grimaud** | 1 | 1 | 0 | 0 |
| **Benjamin Drillat** | 1 | 1 | 8 | 8 |
| **Benoit Nabholz** | 0 | 0 | 5 | 4 |
| **Berenger Remy** | 0 | 0 | 9 | 2 |
| **Blandine Carre** | 0 | 0 | 19 | 6 |
| **Carlos Álvarez-Cros** | 4 | 1 | 1 | 1 |
| **Celine Quelennec** | 0 | 0 | 3 | 2 |
| **Christian Kerbiriou** | 0 | 0 | 7 | 6 |
| **Christian Roesti** | 1 | 1 | 709 | 139 |
| **Clement Lemarchand** | 0 | 0 | 3 | 1 |
| **Clementine Azam** | 0 | 0 | 3 | 3 |
| **Daniel Bizet** | 6 | 1 | 24 | 7 |
| **Daniel Espejo Fraga** | 14 | 5 | 24 | 13 |
| **David Funosas** | 135 | 13 | 116 | 36 |
| **David Sannier** | 175 | 13 | 671 | 97 |
| **Dominik Arend** | 0 | 0 | 579 | 18 |
| **Elodie Massol** | 111 | 7 | 1261 | 59 |
| **Elouan Meyniel** | 0 | 0 | 5 | 4 |
| **Eloïsa Matheu** | 31 | 7 | 219 | 46 |
| **Eric Sardet** | 0 | 0 | 3 | 3 |
| **Evgenia Kovalyova** | 0 | 0 | 8 | 4 |
| **Fabien Sane** | 0 | 0 | 1 | 1 |
| **Fernand Deroussen** | 110 | 9 | 690 | 66 |
| **Florence Matutini** | 0 | 0 | 10 | 6 |
| **Gaëtan Jouvenez** | 5 | 1 | 0 | 0 |
| **Georges Bedrines** | 0 | 0 | 5 | 3 |
| **Ghislain Riou** | 0 | 0 | 23 | 16 |
| **Jakub Burdzicki** | 0 | 0 | 54 | 19 |
| **Joan Estrada Bonell** | 1 | 1 | 10 | 5 |
| **Joan Ventura Linares** | 8 | 5 | 9 | 4 |
| **Jocelyn Fonderflick** | 0 | 0 | 1 | 1 |
| **Joss Deffarges** | 0 | 0 | 31 | 21 |
| **Julien Barataud** | 0 | 0 | 179 | 41 |
| **Julien Cavallo** | 0 | 0 | 2 | 1 |
| **Justine Przybilski** | 0 | 0 | 1 | 1 |
| **Jérome Allain** | 6 | 1 | 0 | 0 |
| **Jérome Sueur** | 719 | 19 | 0 | 0 |
| **K. G.** | 1 | 1 | 0 | 0 |
| **Klaus Alix** | 0 | 0 | 3 | 1 |
| **Laura Martin** | 2 | 2 | 6 | 4 |
| **Laurent Pelozuelo** | 2 | 1 | 163 | 33 |
| **Leslie Campourcy** | 0 | 0 | 2 | 1 |
| **Lucas Roger** | 29 | 6 | 241 | 46 |
| **Lukasz Cudziło** | 0 | 0 | 31 | 25 |
| **Marc Anton** | 1 | 1 | 0 | 0 |
| **Marc Corail** | 0 | 0 | 17 | 12 |
| **Margaux Charra** | 0 | 0 | 44 | 5 |
| **Marie-Lilith Patou** | 0 | 0 | 20 | 5 |
| **Marlene Massouh** | 0 | 0 | 7 | 3 |
| **Marta Celej** | 1 | 1 | 20 | 15 |
| **Mathieu Pélissié** | 7 | 5 | 59 | 37 |
| **Mathieu Sannier** | 0 | 0 | 14 | 11 |
| **Matija Gogala** | 16 | 12 | 0 | 0 |
| **Miguel Domenech Fernández** | 0 | 0 | 10 | 4 |
| **Nicolas Mokuenko** | 3 | 2 | 4 | 4 |
| **Nicolas Vissyrias** | 0 | 0 | 5 | 2 |
| **Pere Pons** | 13 | 7 | 0 | 0 |
| **Piotr Guzik** | 5 | 1 | 194 | 29 |
| **Rafael Carbonell Font** | 119 | 7 | 1245 | 48 |
| **Rafael Tamajón** | 3 | 1 | 0 | 0 |
| **Remi Jullian** | 0 | 0 | 8 | 3 |
| **Romain Riols** | 0 | 0 | 6 | 5 |
| **Roy Kleukers** | 7 | 3 | 183 | 62 |
| **Stéphane Puissant** | 103 | 15 | 1 | 1 |
| **Sylvain Grimaud** | 0 | 0 | 3 | 3 |
| **Szymon Czyzewski** | 0 | 0 | 1 | 1 |
| **Sébastien Merle** | 0 | 0 | 2 | 1 |
| **Tamás Kiss** | 0 | 0 | 5 | 3 |
| **Thomas Armand** | 0 | 0 | 1 | 1 |
| **Tomi Trilar** | 689 | 9 | 0 | 0 |
| **Varvara Vedenina** | 0 | 0 | 6 | 3 |
| **Vincent Milaret** | 6 | 3 | 8 | 6 |
| **Werner Reitmeier** | 0 | 0 | 1 | 1 |
| **Wolfgang Forstmeier** | 3 | 1 | 595 | 121 |
| **Xavier Béjar** | 0 | 0 | 2 | 2 |
| **Yves Bas** | 50 | 6 | 686 | 49 |

**Table S2:** Numerical overview of the contribution of each entomologist to the acoustic dataset

| Category | English label | Original French label |
|---|---|---|
| Anthropophony | Car alarm | Alarme |
| Anthropophony | Camera | Appareil photo |
| Anthropophony | Applause | Applaudissement |
| Anthropophony | Sprinkler | Arroseur |
| Anthropophony | Plane | Avion |
| Anthropophony | Ship | Bâteau |
| Anthropophony | Baby | Bébé |
| Anthropophony | Bip | Bip |
| Anthropophony | Digital bug | Bug informatique |
| Anthropophony | Truck | Camion |
| Anthropophony | Song | Chanson |
| Anthropophony | Collision | Choc |
| Anthropophony | Keys | Clefs |
| Anthropophony | Bell | Cloche |
| Anthropophony | Cowbell | Cloche_v |
| Anthropophony | Scream | Cris |
| Anthropophony | Human movement | Déplacement |
| Anthropophony | Grinder | Disqueuse |
| Anthropophony | Construction machine | Engin_c |
| Anthropophony | Sneeze | Eternuement |
| Anthropophony | Window | Fenêtre |
| Anthropophony | Zip | Fermeture_e |
| Anthropophony | Brake | Frein |
| Anthropophony | Gurgling noise | Gargouillement |
| Anthropophony | Helicopter | Hélicoptère |
| Anthropophony | Undetermined | Indéterminé |
| Anthropophony | Horn | Klaxon |
| Anthropophony | Microphone | Micro |
| Anthropophony | Harvester | Moissonneuse |
| Anthropophony | Engine | Moteur |
| Anthropophony | Motorbike | Moto |
| Anthropophony | Music | Musique |
| Anthropophony | Unidentified sound | Non identifié |
| Anthropophony | Step | Pas |
| Anthropophony | Door | Porte |
| Anthropophony | Scraping | Raclement |
| Anthropophony | Radar | Radar |
| Anthropophony | Radio | Radio |
| Anthropophony | Sniffing | Reniflement |
| Anthropophony | Breath | Respiration |
| Anthropophony | Laughter | Rire |
| Anthropophony | Electric saw | Scie_e |
| Anthropophony | Scooter | Scooter |
| Anthropophony | Whistle | Sifflement |
| Anthropophony | Microphone whistling | Sifflement_m |
| Anthropophony | Alarm | Sirène |
| Anthropophony | Phone ring | Sonnerie |
| Anthropophony | Gunshot | Tir |
| Anthropophony | Coughing | Toux |
| Anthropophony | Tractor | Tracteur |
| Anthropophony | Train | Train |
| Anthropophony | Construction noise | Travaux |
| Anthropophony | Chainsaw | Tronçonneuse |
| Anthropophony | Bike | Vélo |
| Anthropophony | Ventilation | Ventilation |
| Anthropophony | Car | Voiture |
| Anthropophony | Voice | Voix |
| Anthropophony | Shutter | Volet |
| Geophony | Branch | Branche |
| Geophony | Waterfall | Cascade |
| Geophony | Water | Eau |
| Geophony | Leaves | Feuilles |
| Geophony | Water drop | Goutte |
| Geophony | Hail | Grêle |
| Geophony | Creaking | Grincement |
| Geophony | Storm | Orage |
| Geophony | Rain | Pluie |
| Geophony | Water stream | Ruisseau |
| Geophony | Torrent | Torrent |
| Geophony | Wave | Vague |
| Geophony | Wind | Vent |

**Table S3:** List of all labels used to categorize abiotic sounds

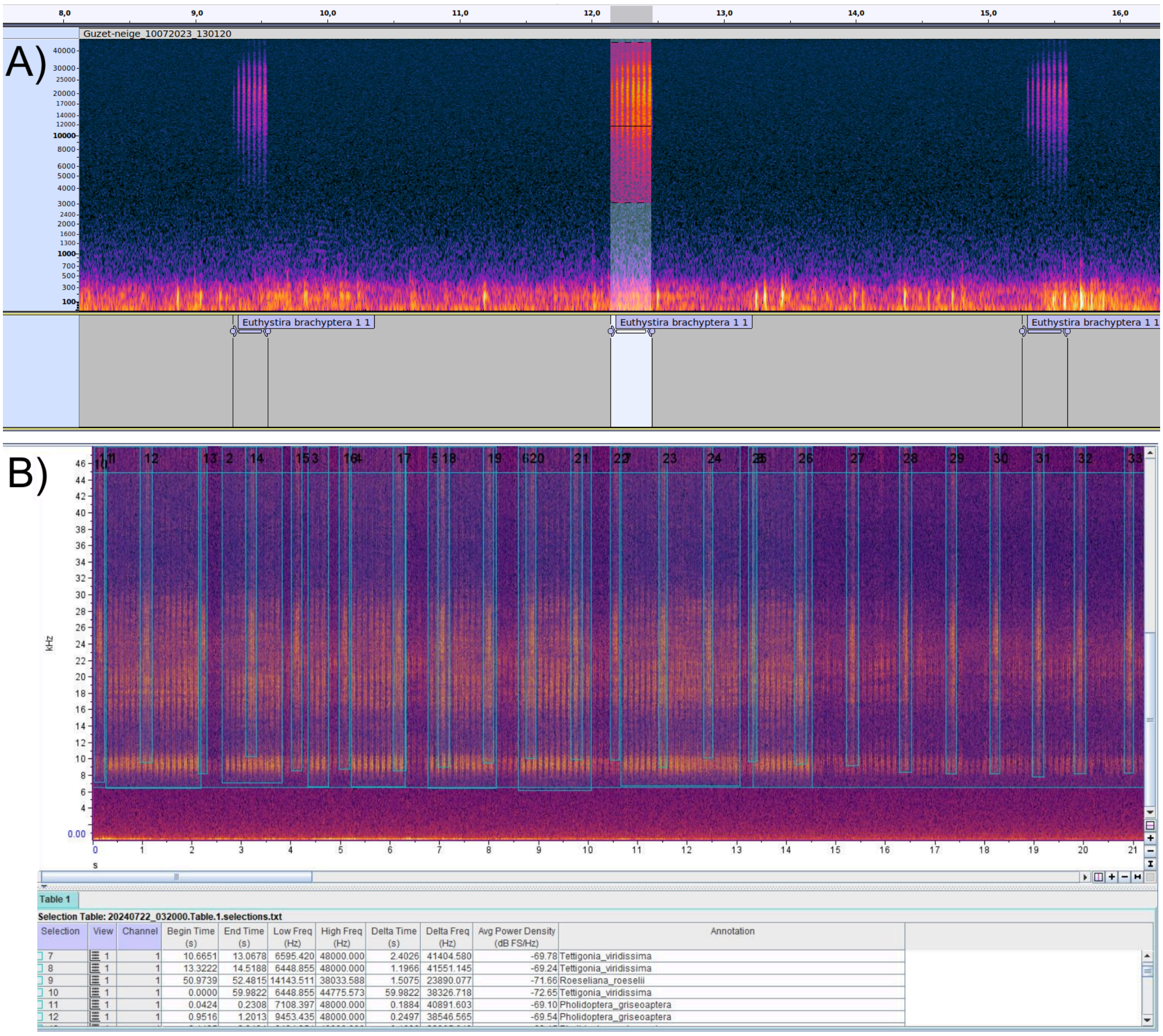

**Figure S1:** Annotations made with A) Audacity and B) Raven Lite 2.0.5: the division on top displays the mel-scale spectrogram of the recording and the division at the bottom displays the manual annotations identifying orthopteran stridulations. All annotations encapsulate a sound both in time and in frequency. Annotations on Raven Lite only consist of the scientific name of the species identified, whereas annotations on Audacity include 2 additional codes: one for the level of confidence in the identification of the species (1 for certain identifications and 0 for uncertain ones) and another for the number of individuals detected in the sound fragment encapsulated by the annotation (1 for 1 individual, 2 for 2 individuals of the same species, 3 for more than 2 individuals of the same species and 4 for multiple individuals of different species).

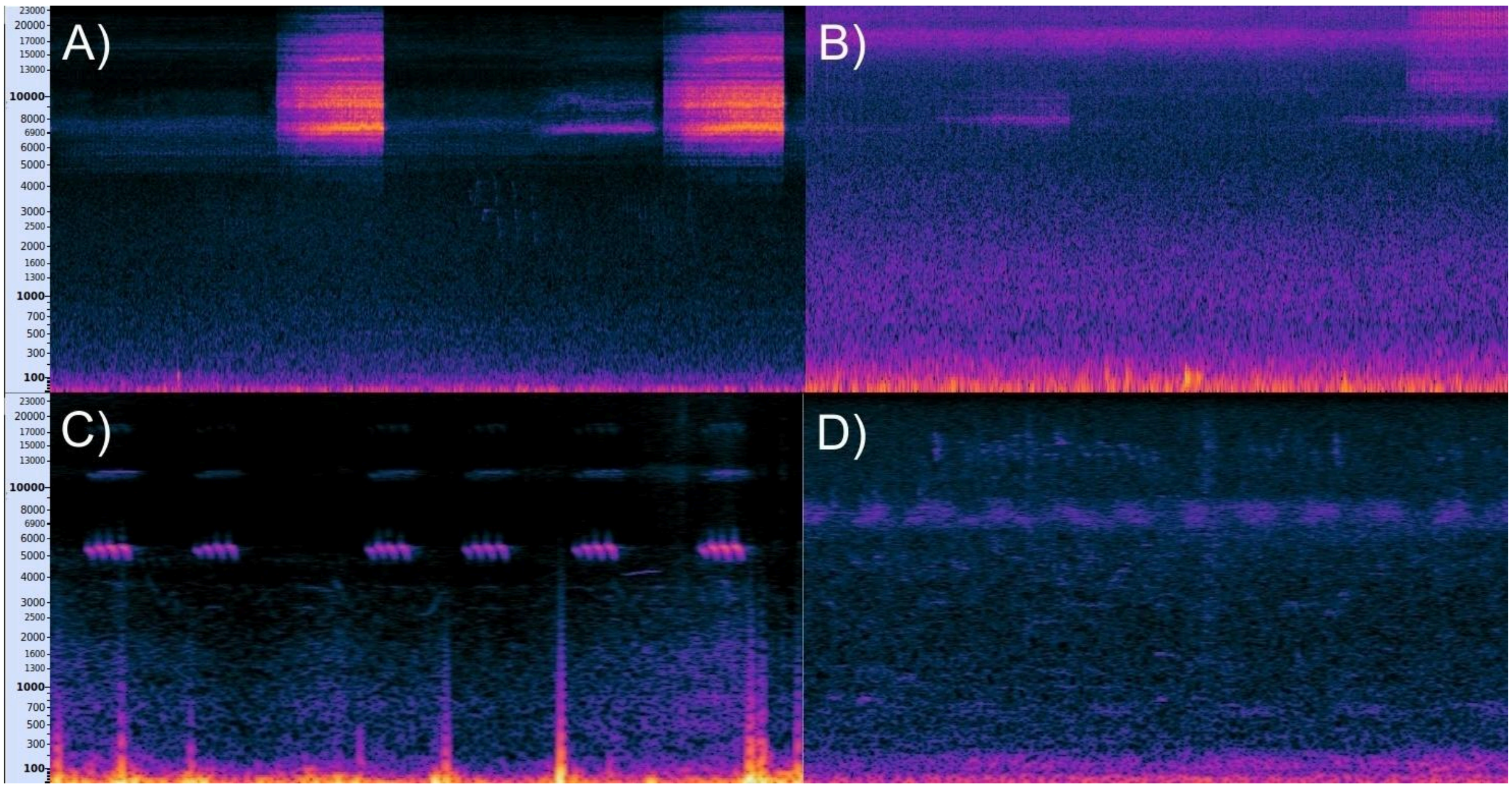

**Figure S2:** Two comparisons of orthopteran songs as they typically appear in focal recordings and in natural soundscapes. The 4 mel-scale spectrograms show songs of A) *Tettigonia cantans* in a focal recording, B) *Tettigonia cantans* in a soundscape, C) *Gryllus campestris* in a focal recording, and D) *Gryllus campestris* in a soundscape.